\documentclass[aps,pra,twocolumn,longbibliography,superscriptaddress,floatfix]{revtex4-1}
\usepackage{amsfonts}
\usepackage{mathtools}
\usepackage{graphicx}
\usepackage{epsfig}
\usepackage{dcolumn}
\usepackage{bm}
\usepackage{amsmath}
\usepackage{graphicx}
\usepackage[latin1]{inputenc}
\usepackage{ulem}
\usepackage{epstopdf}
\usepackage{subfigure}
\usepackage{color}
\usepackage{amsthm}
\usepackage{newlfont}
\usepackage{graphicx}
\usepackage{epstopdf}
\usepackage{appendix}

\begin{document}
 \title{How close Are Integrable and Non-integrable Models: \\
 A Parametric Case Study Based on the Salerno Model}
\author{Thudiyangal Mithun}
\affiliation{Department of Mathematics and Statistics, University of Massachusetts, Amherst MA 01003-4515, USA} 
\author{Aleksandra Maluckov}
\affiliation{COHERENCE, Vin\v ca Institute of Nuclear Sciences, National Institute of the Republic of Serbia, University of Belgrade, P. O. B. 522, 11001 Belgrade, Serbia}
\author{Ana Man\v ci\'c}
\affiliation{COHERENCE, Dept. of Physics, Faculty of Sciences and Mathematics, University of Ni\v s, P.O.B. 224, 18000 Ni\v s, Serbia}
\author{Avinash Khare}
\affiliation{Department of Physics, Savitribai Phule Pune University, Pune 411007, India}
\author{Panayotis G. Kevrekidis}
\affiliation{Department of Mathematics and Statistics, University of Massachusetts, Amherst MA 01003-4515, USA} 

\begin{abstract}
In the present work we revisit the Salerno model as a prototypical system that interpolates between a well-known integrable system (the Ablowitz-Ladik lattice) and an experimentally tractable non-integrable one (the discrete nonlinear Schr{\"o}dinger model). The question we ask is: for ``generic'' initial data, how close are the integrable
to the non-integrable models? Our more precise formulation of this question is: how well is the constancy of formerly conserved quantities 
preserved in the non-integrable case? Upon examining this, we find
that even slight deviations from integrability can be 
sensitively felt by measuring these formerly conserved quantities
in the case of the Salerno model. However, given that the knowledge
of these quantities requires a deep physical and mathematical 
analysis
of the system, we seek a more ``generic'' diagnostic towards a manifestation of integrability breaking. We argue, based on 
our Salerno model computations, that the full spectrum of Lyapunov exponents could be a sensitive diagnostic to that effect.
\end{abstract}

\maketitle
\pagestyle{myheadings}

\section{Introduction}

The topic of nonlinear dynamical lattices and energy 
localization in them has been prevalent in a large
array of studies over the past few decades~\cite{FPUreview,pgk:2011}.
Indeed, since the proposal of intrinsic localized modes in
anharmonic crystals~\cite{ST,Page}, there has been 
an ever expanding range of disciplines where relevant
states and their implications are being identified,
explored and dynamically exploited~\cite{Flach:2008}.
Among the numerous associated examples, one can list
arrays of waveguides in nonlinear optics~\cite{moti},
Bose-Einstein condensates in optical lattices~\cite{Morsch},
manipulation of localization in micromechanical oscillator
arrays~\cite{sievers}, granular crystals in 
materials science~\cite{yuli_book,granularBook}, 
lattices of electrical circuits~\cite{remoissenet},
and many others including layered antiferromagnetic crystals~\cite{lars3,lars4}, Josephson-junction ladders~\cite{alex,alex2}, or 
dynamical models of the 
DNA double strand \cite{Peybi}.

In many of these works, part of the emphasis has
been on localization and nonlinear wave structures~\cite{Aubry06,Flach:2008,kev2009,pgk:2011}.
Important associated questions involve the existence,
dynamical stability and nonlinear dynamics of the relevant
waveforms. A parallel line of activity that has also been
central from early on has been that of potential long-time ergodicity of the nonlinear lattice dynamical 
systems~\cite{Fermi:1955,FPUreview}. In the latter, there
have been significant developments in recent times,
where computational resources have enabled far longer 
time simulations of different classes of such
systems~\cite{Danieli:2017,Danieli:2019,Mithun:2019} and
the development of novel systems that are more straightforward
to simulate over long times~\cite{merab}. Interstingly, the
birth of the scientific field examining 
nonlinear wave (solitonic) structures 
has been strongly connected with such ergodicity-related quests~\cite{Zabusky:1965,Porter:2009}.

The concept of integrability~\cite{AS81,ablowitz2} is one
that is central to both of the above directions of study.
On the one hand, the development of the inverse scattering
transform and the identification of solitonic structures for
a number of these equations has been a key development in
nonlinear wave dynamics~\cite{AS81,ablowitz2}, while on the
other hand, the infinite conservation laws and associated
constraints that such systems impose on the dynamics have
significant bearings on the ability of the system to 
explore its phase space. Moreover, often integrability
has been a ``helpful hand'' towards trying to understand
the dynamics of weakly non-integrable systems through
approaches involving perturbation theory~\cite{RevModPhys.61.763}.
Here, often an effective adiabaticity assumption is implied, i.e., that 
the structures of the integrable (or analytically tractable)
limit are preserved but their features (e.g., amplitude, width,
speed, etc.) are modified and dynamically driven by the 
non-integrable perturbations imposed. Indeed, this proximity
has been recently also of substantial mathematical interest
through, e.g., the works of~\cite{karch1,karch2}.

In the present work, it is our intention to return to the
exploration of this topic of the effective proximity of
integrable and weakly non-integrable systems. Indeed, 
we leverage here a different perspective from those
of works such as~\cite{karch1,karch2} which focus on the
(small) amplitude of the solution to gauge the relevant
proximity. Rather, we deploy a comparison on the basis
of conservation laws of the original integrable system
(see also the work of~\cite{PhysRevE.68.056603}). Our
aim is to explore more broadly the phase space of
the lattice dynamical system and its constraints as we depart from the
integrable limit. As our platform of choice, we will
utilize the well-known so-called Salerno model~\cite{salerno1992quantum}, given its natural
interpolation between the well-established
integrable variant of the nonlinear Schr{\"o}dinger
equation (the so-called Ablowitz-Ladik (AL) limit)~\cite{ablowitz1976nonlinear,ablowitz_prinari_trubatch_2003} and the non-integrable so-called DNLS
(discrete nonlinear Schr{\"o}dinger) equation~\cite{kev2009}.
The advantage of this system is the availability of 
a homotopic parameter interpolating between these models
and allowing us to explore the departure from the integrable
limit. 

Our tool of choice will be the usage of 
conservation laws of the AL limit initially.
We will explore how ``sensitive'' these are as
probes of the breaking of integrability. We will
find that indeed ``former conservation laws'' will
be very sensitive to departures from the relevant limit.
However, a disadvantage of this approach is that it
requires a deep mathematical or physical (or both) knowledge
of the concrete features of the system at hand. In that
light, it is desirable to have a more general toolbox
that is somewhat ``system independent'' in order to
(sensitively) probe such departures from the integrable 
limit. In that vein, we explore the maximal Lyapunov
exponent and, indeed, the full Lyapunov spectrum of
the system of interest that can be generally 
computed~\cite{benettin:1980a,benettin:1980b}. We find
that this represents a very efficient tool for detecting
the number of available conservation laws and hence
integrability of the system, indeed one that we
expect in the future to be amenable to efficient
computation, e.g., via machine-learning techniques.

Our presentation will be structured as follows.
In section II, we present the model of interest
and its associated conservation laws that we will
probe both in the integrable limit and systematically
as we depart from that limit. In section III,
we present our results for the corresponding
conservation laws and their long-time dynamics.
In section IV, we discuss the computation of the
Lyapunov exponent spectrum, both as regards the
maximal Lyapunov exponent and as regards the
full spectrum and present associated numerical
results. In section V we summarize our findings
and present our conclusions, as well as a number
of directions for future study.

\section{Model description}

The equation that we will consider in the present study
involves the well-established Salerno model~\cite{salerno1992quantum}, which interpolates
between the AL and the DNLS limits. The relevant
dynamical equation reads:
\begin{eqnarray}
\begin{split}
\label{eq1a}
i\frac{d\psi_n}{dt}&=
(1+\mu |\psi_n|^2)(\psi_{n+1}+\psi_{n-1}){+}\gamma |\psi_n|^2\psi_{n}.\label{equa1}
\end{split}
\end{eqnarray}
This system has been a natural playground 
for the usage of perturbation theory methods
off of the integrable limit~\cite{PhysRevE.53.4131},
for the examination of the delicate issue of mobility
in lattice dynamical systems~\cite{GOMEZGARDENES2004213},
for the exploration of collisions~\cite{PhysRevE.68.056603},
and for the analysis of statistical mechanical 
properties of nonlinear lattices~\cite{Mithun_Salerno2021},
among many others.

The AL model is well-known to be integrable via
the inverse scattering transform~\cite{ablowitz_prinari_trubatch_2003}.
This implies the existence of an infinite number
of conserved quantities
 considered, e.g., in the work of~\cite{cassidy2010chaos}, while the non-integrable DNLS limit is characterized 
 solely by two integrals of motion, namely the energy
 and the (squared) $l^2$ norm of the field.
 Indeed, the Salerno model inherits these two conservation
 laws. 
More specifically, regardless of the limits,  Eq.~\eqref{equa1} can be characterized by two
conserved quantities: the (squared) norm $\mathcal{A}$ and the Hamiltonian $\mathcal{H}$, i.e., the energy
of the model~\cite{cai1994moving,PhysRevE.53.4131} in the
form:
\begin{equation}
\begin{split}
\mathcal{A}&=\sum_{n=1}^N \mathcal{A}_n,~~\mathcal{A}_n =\frac{1}{\mu} \ln{|1+\mu|\psi_n|^2|} \\
\mathcal{H}&=\sum_{n=1}^N\Big[-\frac{\gamma}{\mu} \mathcal{A}_n+\psi_n\psi_{n+1}^*+\psi_n^*\psi_{n+1}+\frac{\gamma}{\mu} |\psi_n|^2 \Big],
\label{eq3}
\end{split}
\end{equation}
where $N$ is the total number of lattice nodes
and periodic boundary conditions are used. 
Notice that the latter will be an important point,
especially when we consider finite, small-size
lattices, as integrability of the AL model is
preserved in the case of periodic boundary conditions,
although other types of integrable boundary conditions
may also exist~\cite{CAUDRELIER2019114720}.
It is also relevant to point out that in the DNLS
limit of $\mu \rightarrow 0$, application of l'Hospital
(or a Taylor expansion in $\mu$) leads to the 
first conserved quantity turning into 
the (squared) $l^2$ norm.

The dynamical equations of the Salerno model in the form
of Eqs. \eqref{equa1} can be derived from the Hamiltonian $\mathcal{H}$ according to:
\begin{equation}
\frac{d\psi_n}{dt}=\{\mathcal{H},\psi_n\}.
\end{equation}
with respect to the canonically conjugated pairs of variables $\psi_n$ and $i\psi_n^*$ defining the deformed Poisson brackets \cite{Cai:1994}
\begin{equation}
\label{eq4}
\{\psi_n,\psi_m^*\}=i(1+\mu|\psi_n|^2)\delta_{nm},~~
\{\psi_n,\psi_m\}=\{\psi_n^*,\psi_m^*\}=0.
\end{equation}

Among the infinite conservation laws of the AL limit,
 the two that we will focus on observing here are~\cite{cassidy2010chaos}:
\begin{eqnarray}
 C_1&=& -\mu \sum_n \psi^{\ast}_n\psi_{n-1}\\ \nonumber
 C_2&=& -\mu \sum_n \psi^{\ast}_n\psi_{n-2}(1+\mu |\psi_{n-1}|^2+\frac{1}{2}(\psi^{\ast}_n\psi_{n-1})^2).
 \label{eq:conserve}
\end{eqnarray}
These will be our monitored quantities (that, as we
will see, will be quite informative) and which, 
in the following, will be denoted as moments. In general they are complex quantities. Therefore we considered their real and the imaginary parts, as well as the corresponding modulus. 

 It is relevant to point out that $C_1$ is often thought
 of as a discrete version of the momentum, yet $C_2$ 
 does not have an immediate interpretation at a physical
 level.
  As an additional relevant remark, the moment $C_1$ 
  is not sufficient in order to showcase the integrable
  limit here, as, for instance, it is still conserved
  as a quantity in the linear limit of $\mu=\gamma=0$ (not considered
  in detail herein). On the other hand, the quantity $C_2$ is strictly conserved in the AL integrable case only and
  hence the combination of these two moments should
  be able to provide us a clearer signature associated with
  the integrable limit in what follows.

\begin{figure}[!htbp]
		\includegraphics[angle=0, width=0.9\linewidth]{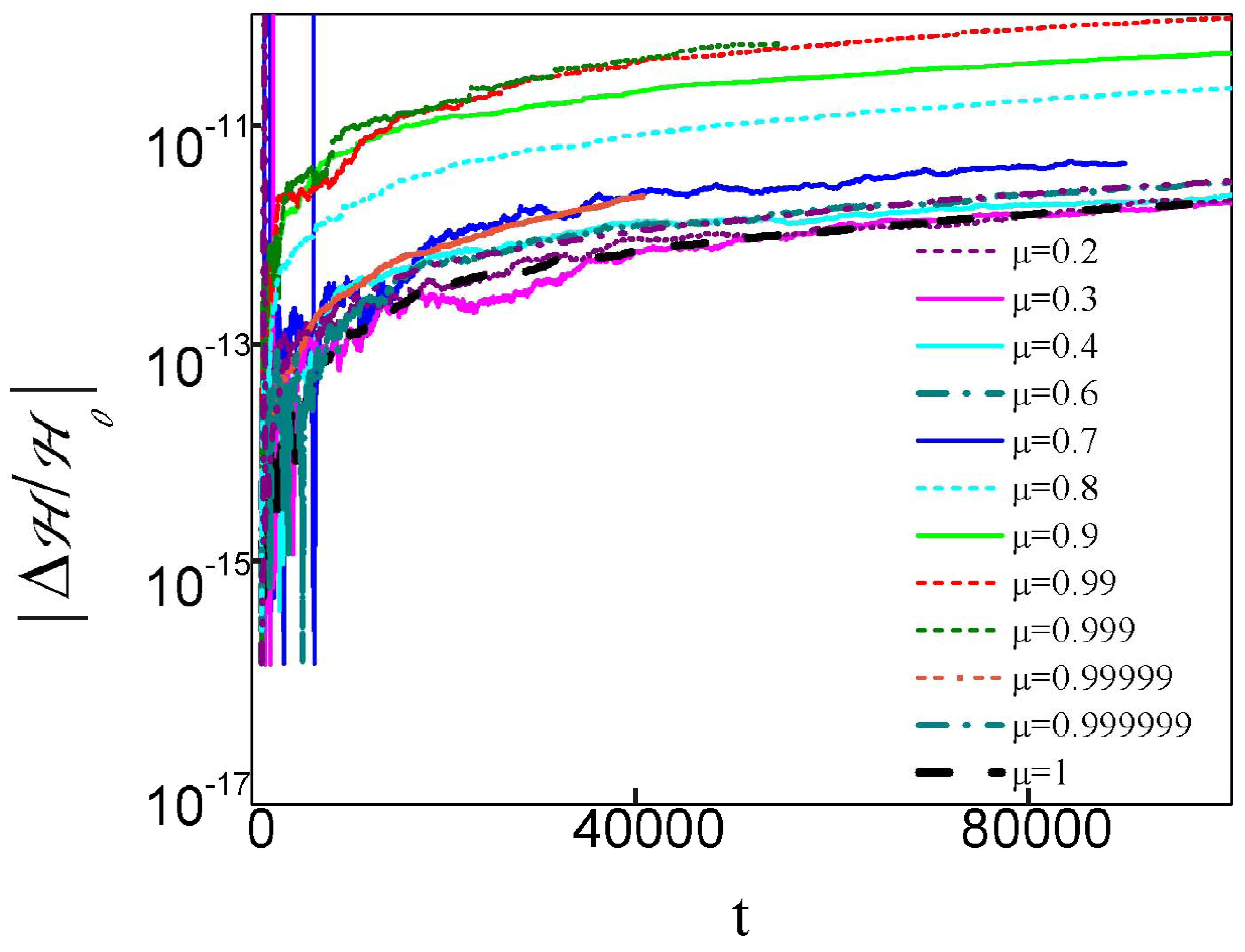}
		\caption{The relative error of energy $\Delta \mathcal{H}/\mathcal{H}_0$, $\Delta \mathcal{H}=\mathcal{H}(t)-\mathcal{H}_0$, for different $\mu$ vs time for 
		an end time 
		of $100000$. A numerical Runge-Kutta (RK) procedure with adaptive step is applied. The values of $\mu$ are shown in the figure, while $N=100$. 
		During the calculations in order for the code to run in the area above $\mu=0.9$ the initial time step, $dt$, was changed from $0.0001$ to $0.000001$. Note that this is only the initial parameter for the adaptive step numerical method. 
		Some of the relevant computations have been stopped
		at shorter times (for reasons that have to do with
		error tolerances applied to these long runs).}
\label{fig:error}
\end{figure}

\begin{figure*}[!htbp]
		\includegraphics[angle=0, width=\linewidth]{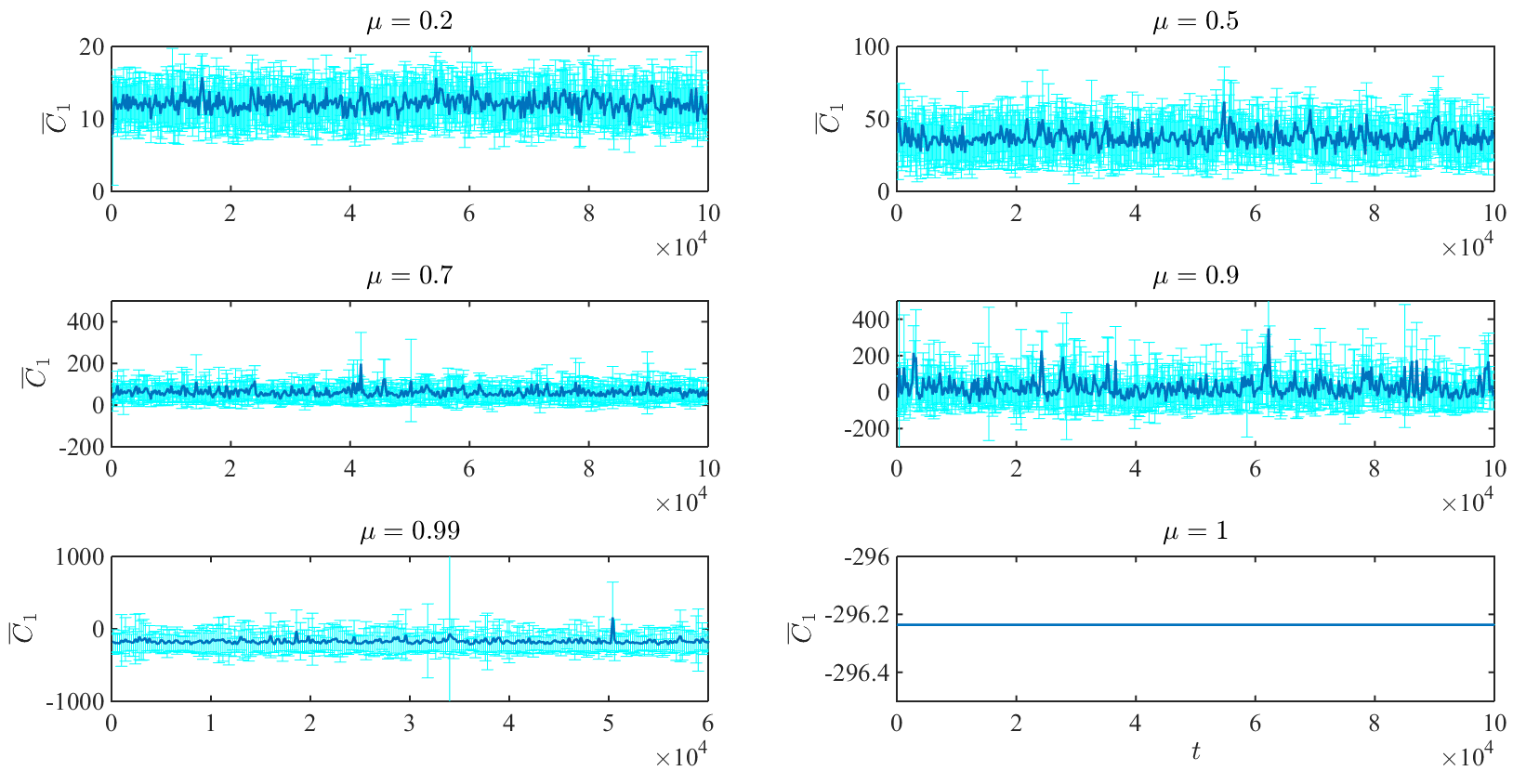}
			\includegraphics[angle=0, width=\linewidth]{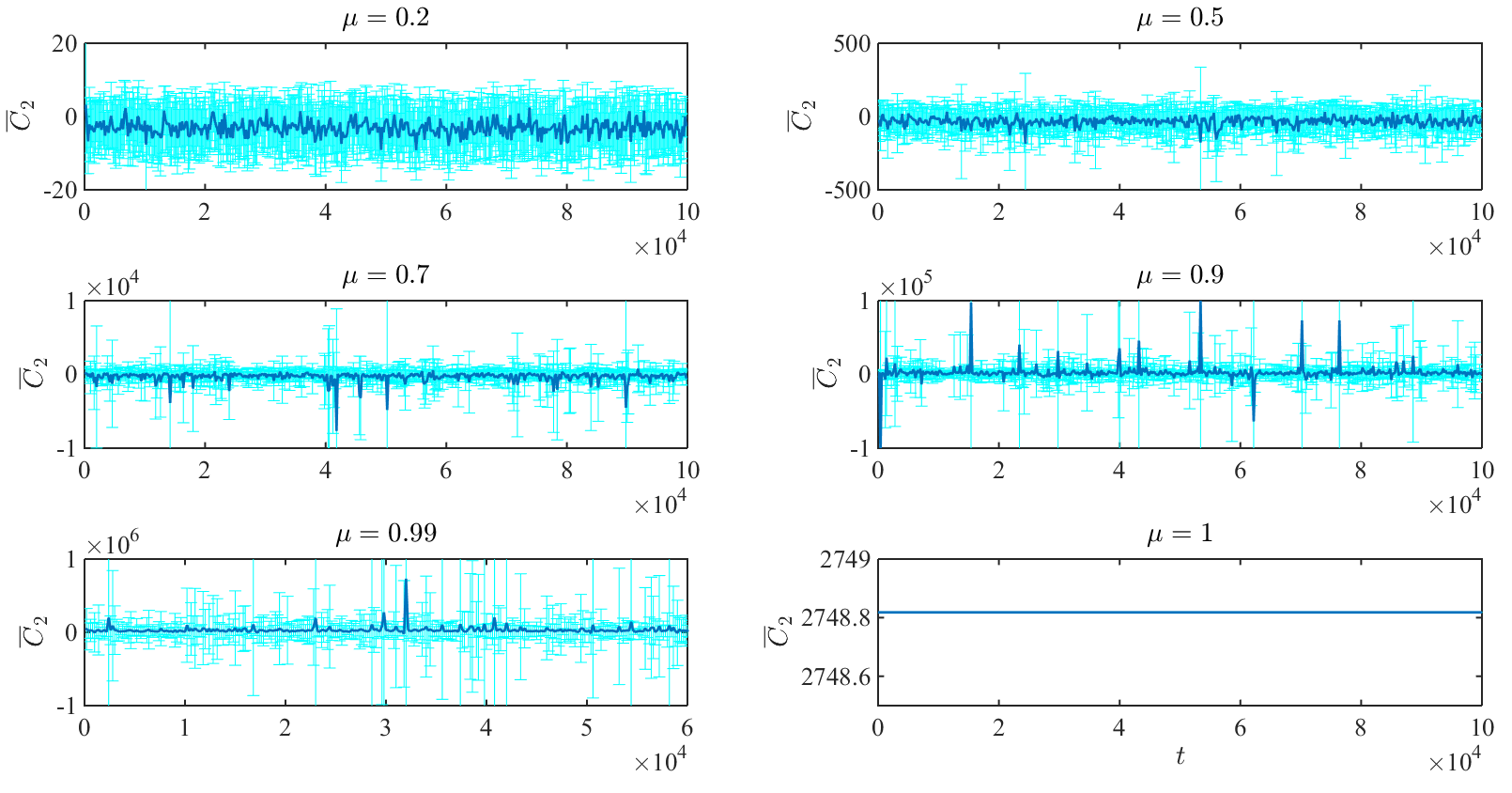}
	
	\caption{The averaged over the successive $\Delta t=100$a.u.
	(a.u. stands for arbitrary time units) real parts of the complex moments,  $\overline{C}_1$ and $\overline{C}_2$ (i.e., the quantities that are 
	conserved in the AL limit but not away from it) vs. $\mu$. The moments are numerically calculated at each selected time step using the equation of motion (\ref{equa1}). Initially,
	at $t=0$, $N=100$ lattice nodes at each $\mu$ are excited by the plane wave with parameters $a\equiv \mathcal{A}/N=1.5$ and 
	$h=\mathcal{H}/N= 3$  to which {a complex random perturbation in space is added by means of a numerical uniform random generator. The set of random numbers from the interval between $(-0.5,0.5)$ is used while the strength of the corresponding perturbation is $0.001$.}  
Bars denote the standard deviation around the mean value taken over a time interval of $t=100$ along the whole propagation time. The values of $\mu$ are specified in the plot.}
	\label{fig:C1C2evolve}
\end{figure*}

\begin{figure*}[!htbp]
		\includegraphics[angle=0, width=\linewidth]{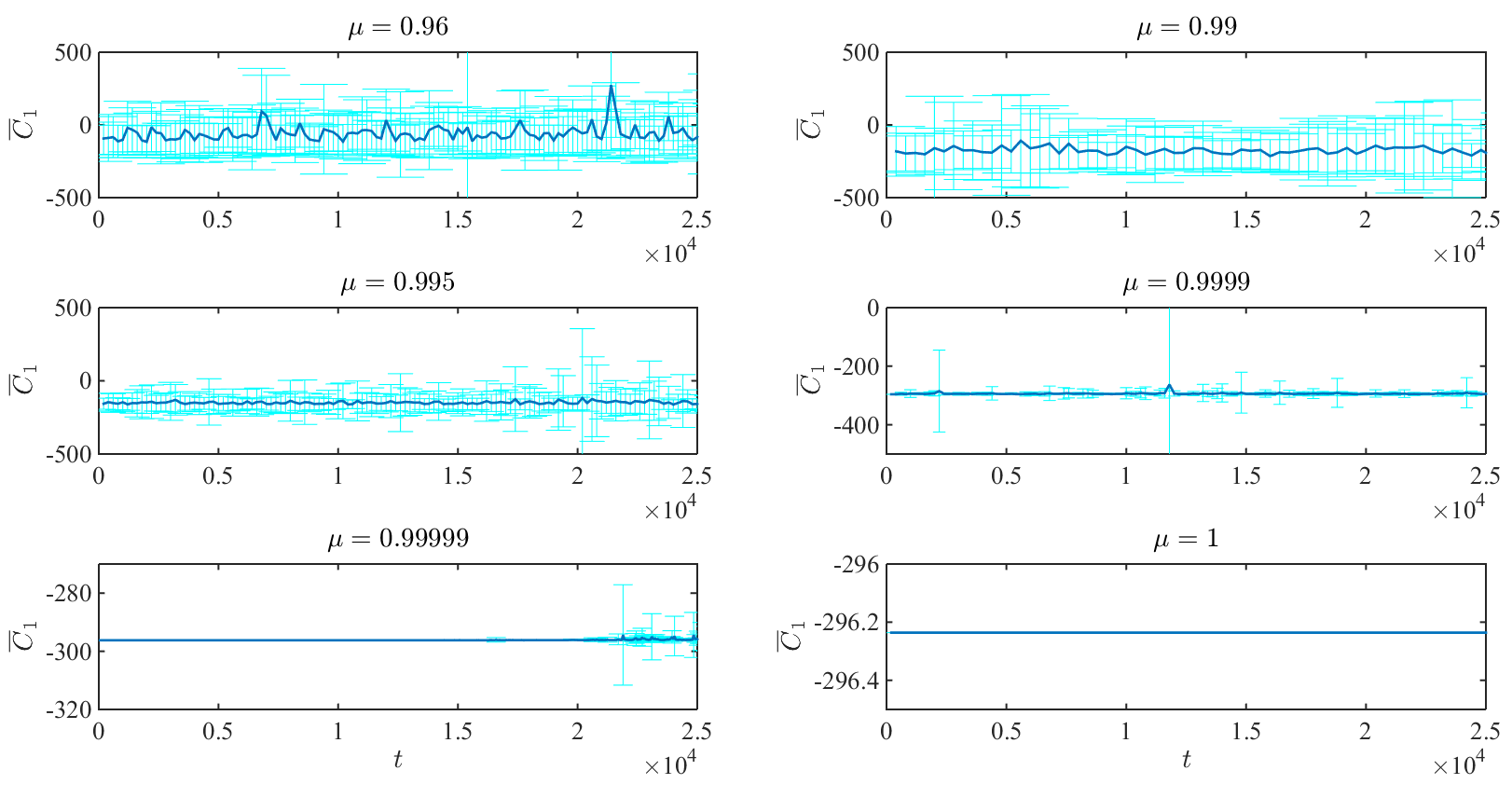}
			\includegraphics[angle=0, width=\linewidth]{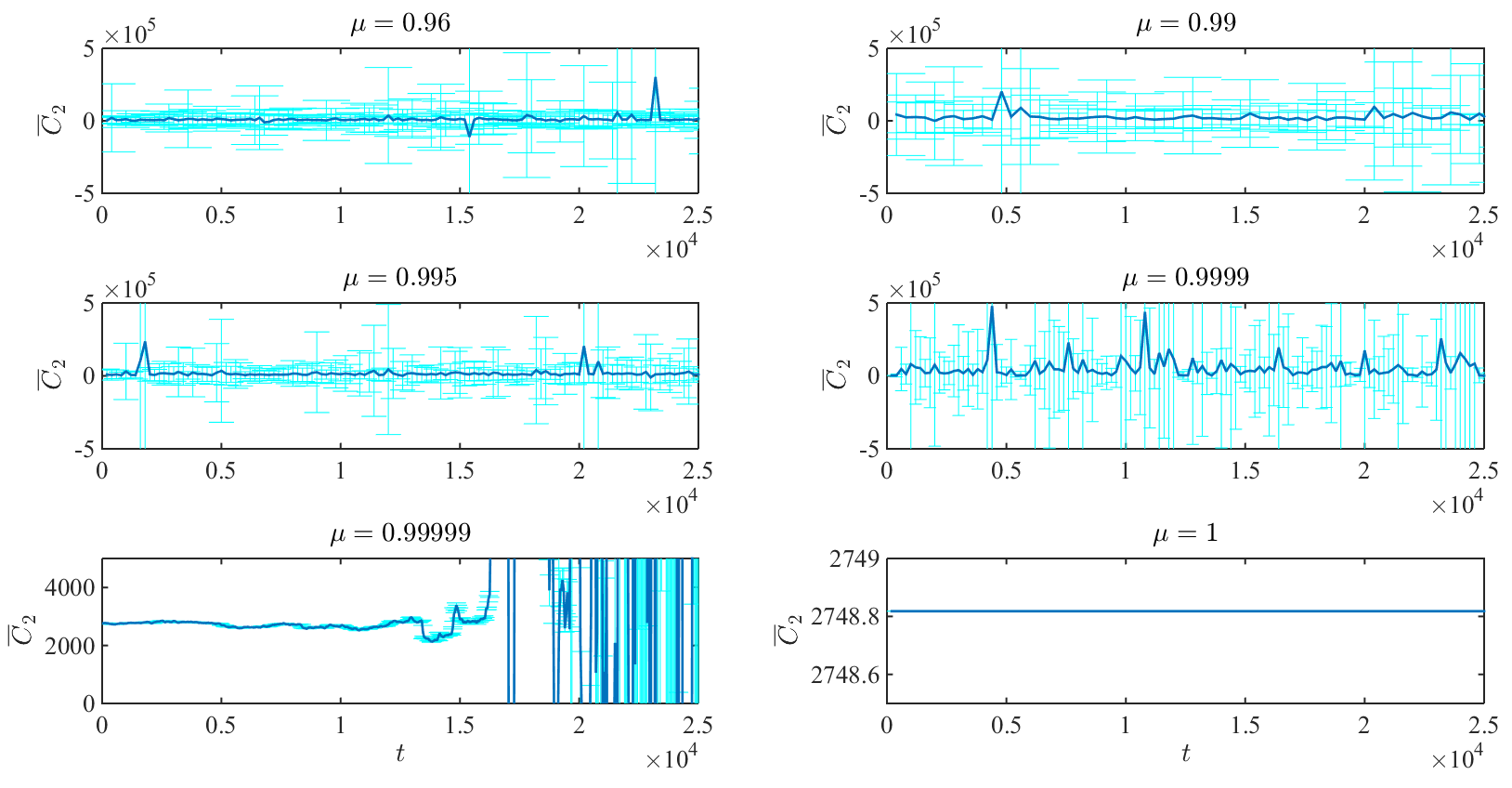}
	
	\caption{ Here we separately show the behaviour of the same moments as in Fig.~\ref{fig:C1C2evolve}, namely
	 $\overline{C}_1$ (first three rows) and $\overline{C}_2$ (last three rows) for $\mu$ very close to the integrability limit. The
	 features and initialization of these runs are similar
	 to the ones of the previous figure.}
	 \label{sharp2}
\end{figure*}

\begin{figure*}[!htbp]
	\includegraphics[width=\linewidth]{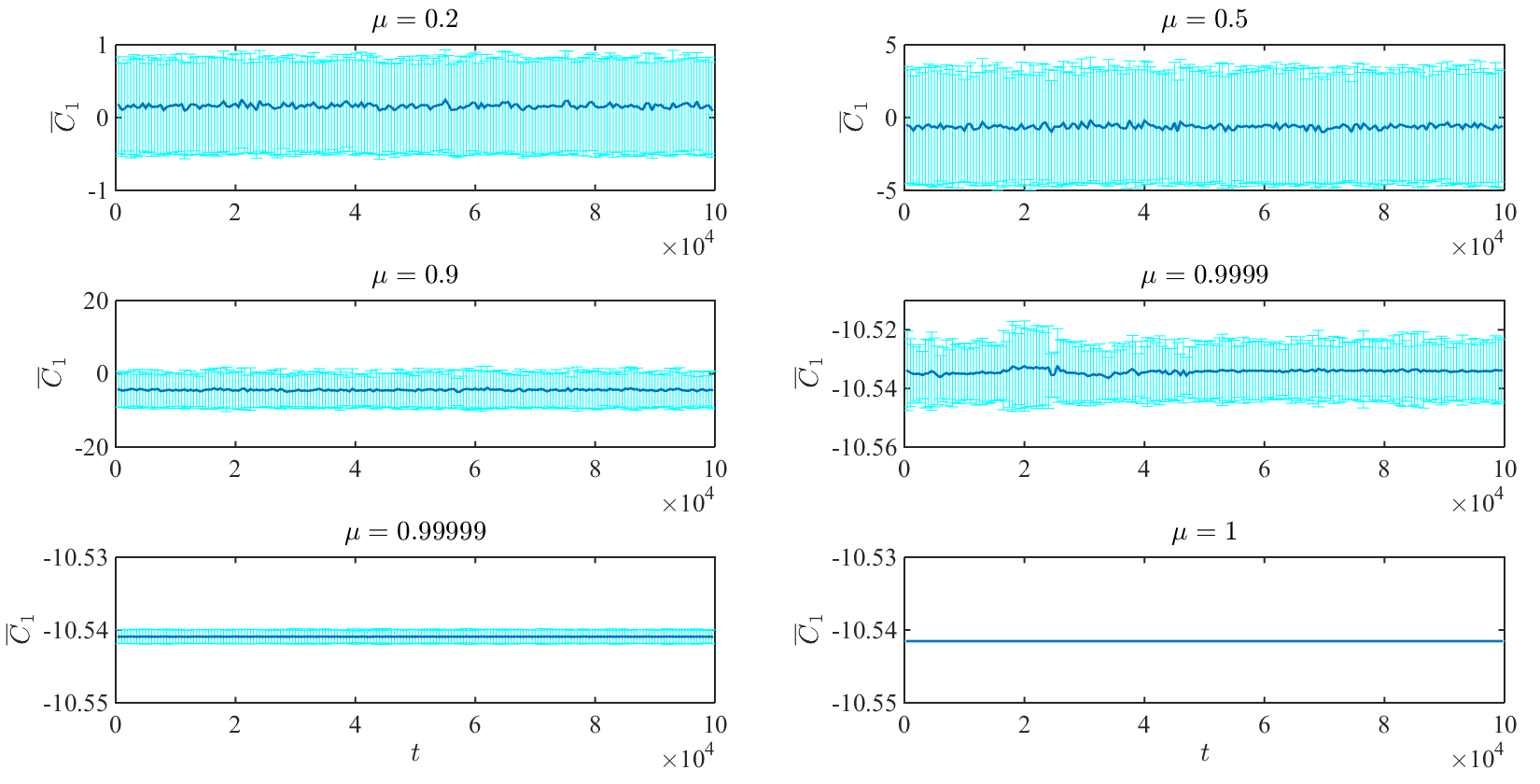}
	\includegraphics[width=\linewidth]{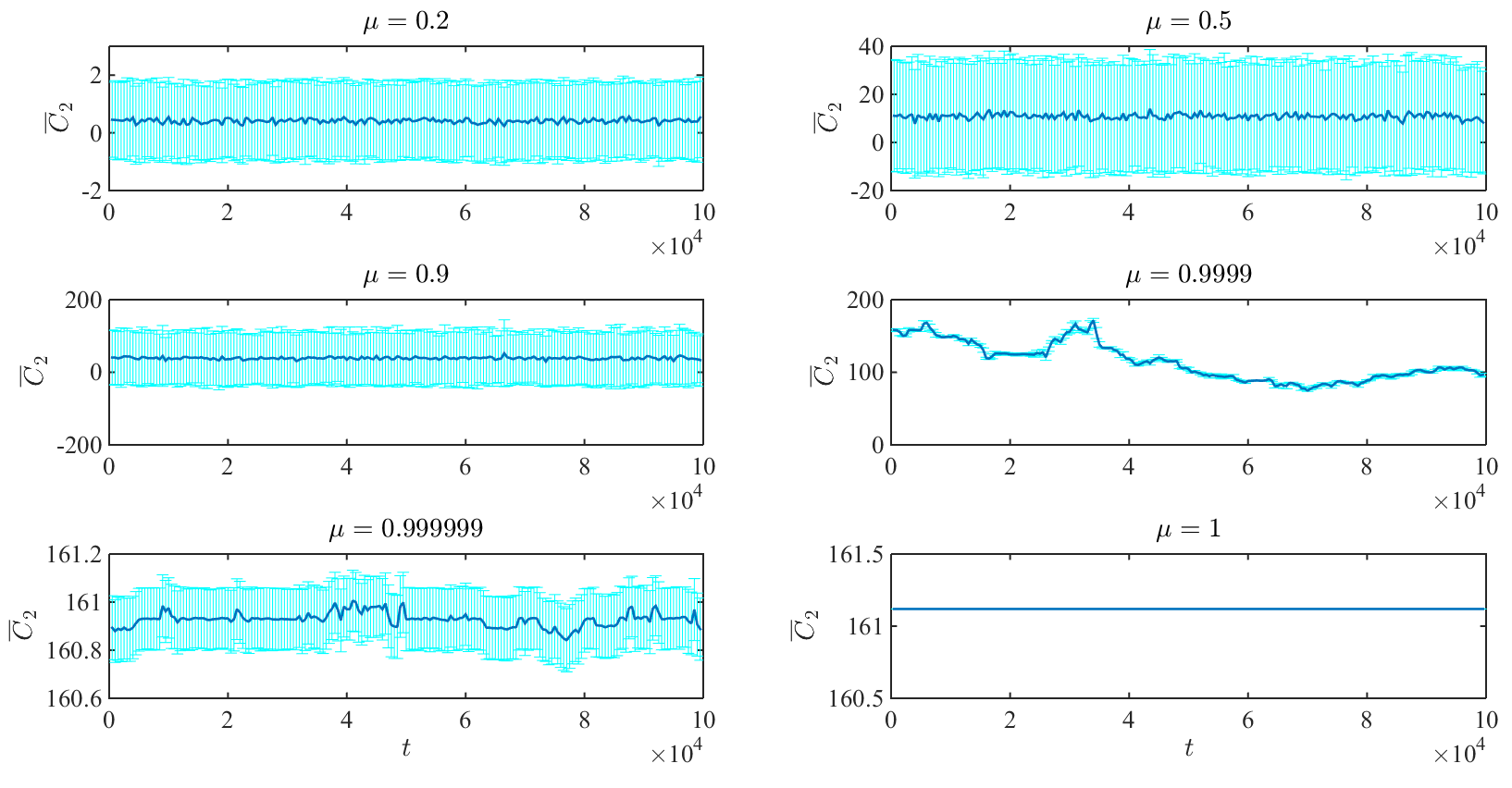}
	\caption{{Here, we present the same moment information
	as in the previous Figure, but for $N=5$. 
	}} \label{moment5}
	
\end{figure*}

\begin{figure}[hh]
	\includegraphics[width=\columnwidth]{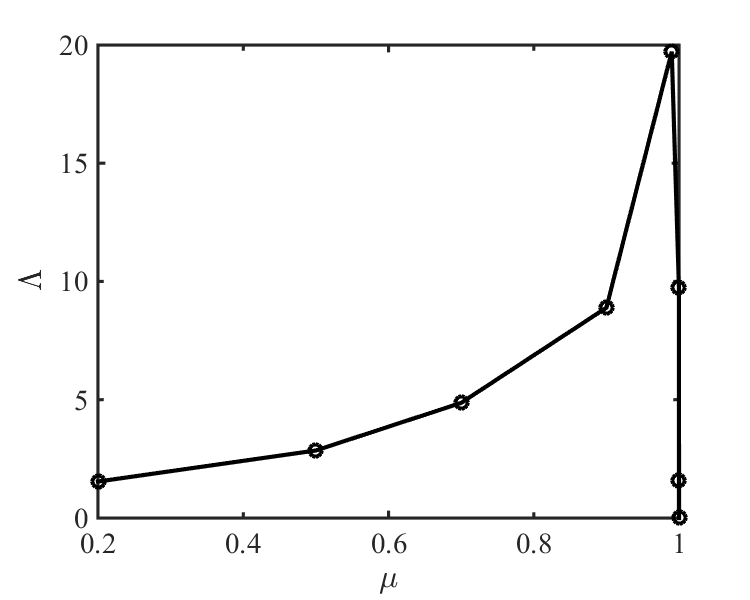}
	\caption{The mLCE defined by Eq.~(\ref{mlce}) for different values of $\mu$ for  fixed $N=100$; $a=1.5$ and $h=3$.
	Notice the positive value thereof except when we
	approach the integrable limit $\mu=1$, when it tends to $0$.}  \label{mLCE100}
\end{figure}

\begin{figure*}[!htbp]
	\includegraphics[width=18cm]{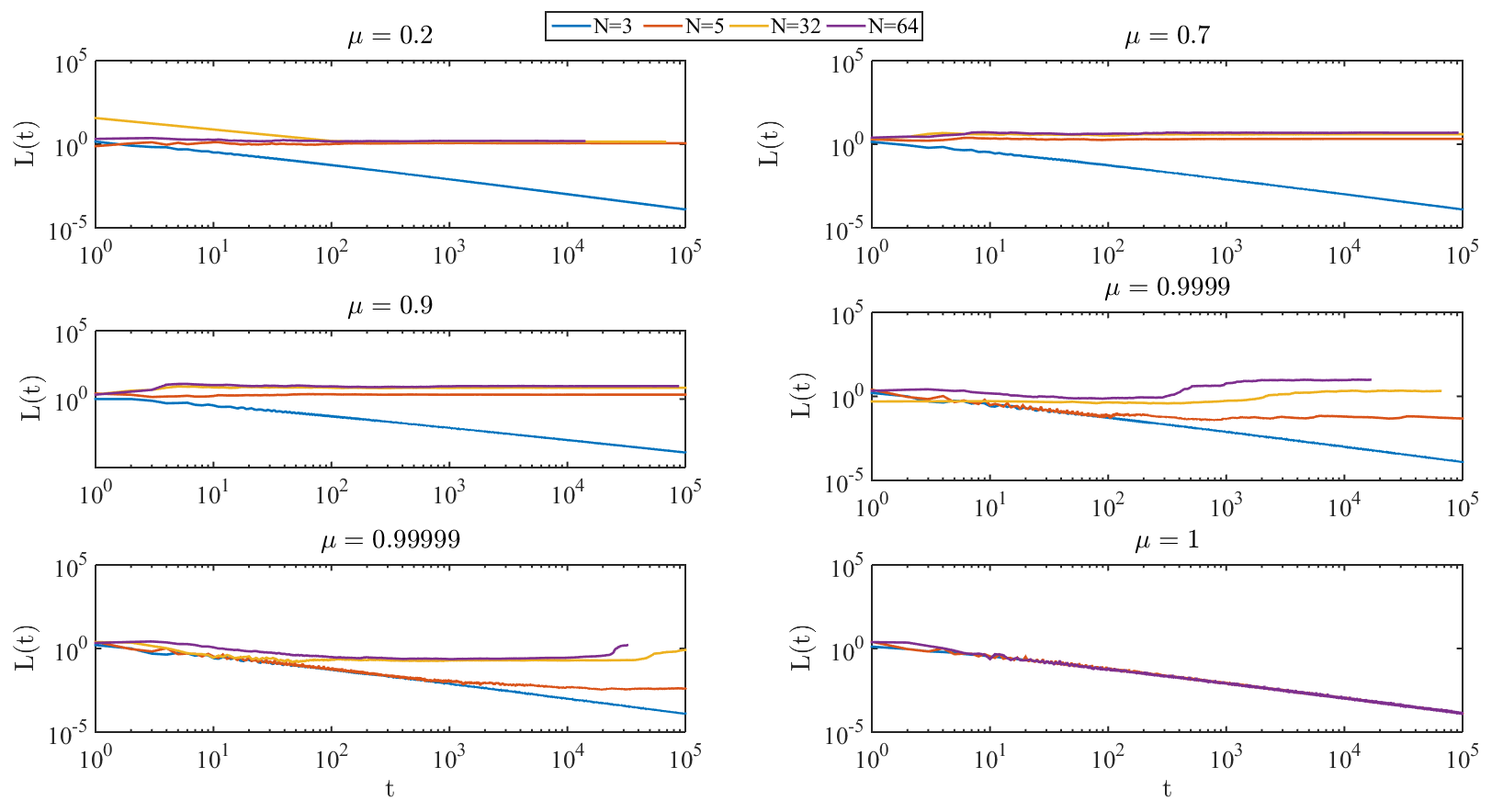}
	\caption{{$L(t)$ in log-log scale for different values of the lattice size $N$ and of the nonlinearity parameter $\mu$ obtained from Eq.~(\ref{mlce}).}} \label{LiapunC1}
\end{figure*}

\begin{figure}[!htbp]
	\includegraphics[width=\columnwidth]{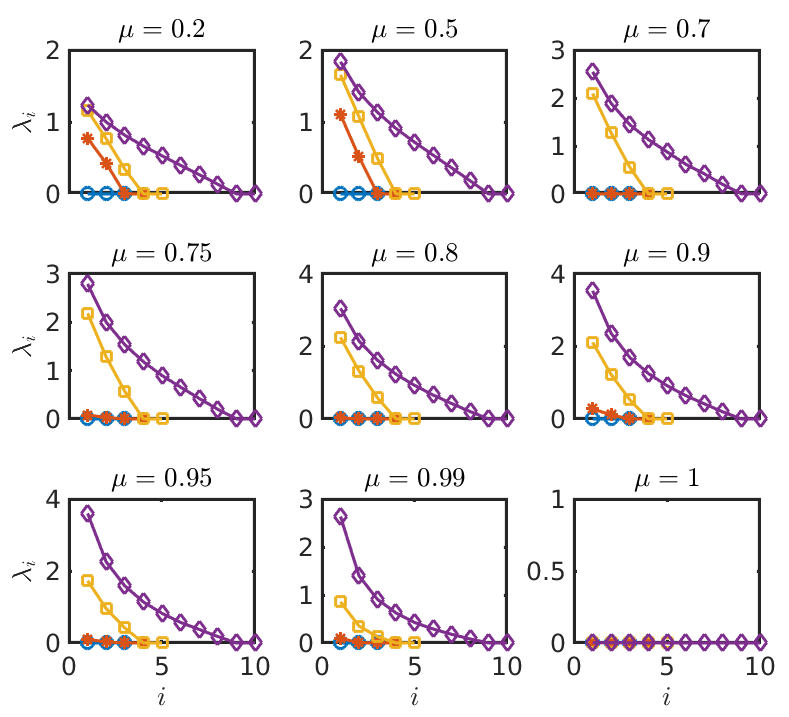}
	\caption{The positive part of the Lyapunov spectrum, i.e., the positive 1D Lyapunov exponents $\lambda_i$  for different values of $\mu$. Here, the number of spectrum components is associated to the x-axis and for each case $i_{total}=N$. The circles, star, square and diamond symbols represent $N=3$, $N=4$, $N=5$ and $N=10$ respectively. {The Lyapunov spectrum (LS) of the complex wave-function associated with the dynamical evolution of the Salerno lattice with $N$ sites, Eq.~(\ref{equa1}), consists of $2N$ one-dimensional (1D) Lyapunov coefficients $\lambda_i, \, i=1,...,2N$. 
	Here we consider the $N$ dependence of the LS 
	near and at the integrability limit. 
We derived the Lyapunov spectrum adopting the procedure described in \cite{lichtenberg1992regular} which is based on the approach of \cite{benettin:1980a,benettin:1980b}.  Here we present the positive $N$ Lyapunov coefficients, $\lambda_{i},~~(i=,1,2,...,N)$}.} \label{fig:lspectrum}
\end{figure}

\section{Computation of Conservation Laws}
In presenting our results,
we start with the aforementioned conserved quantities of Eq.~\eqref{equa1}. Here we gradually deviate
from the AL limit of $\mu=1$, while numerically solving Eq.~\ref{equa1} using an explicit Runge-Kutta algorithm 
of adaptable order~\cite{freelyDOP853,hairer1993solving,mine-01-03-447}. 
In the process, we trace the evolution of the relative norm and energy error, $|\frac{\mathcal{A}(t)-\mathcal{A}(0)}{\mathcal{A}(0)}|$ and  $|\frac{\mathcal{H}(t)-\mathcal{H}(0)}{\mathcal{H}(0)}|$, respectively and show a typical example thereof
in Fig.~\ref{fig:error}. It is relevant to point out here
that this is typically less well-conserved than the
$l^2$ norm (due to the effective presence of a discrete
analogue of the gradient). Nevertheless, it can be
seen that the energy is extremely well conserved with
a relative error below $1e^{-10}$ for the simulation horizons
reported herein.

Fig.~\ref{fig:C1C2evolve} shows the time development of the quantities $C_1$ and $C_2$ for different values of $\mu$. 
The important feature to observe in these evolution simulations
is that the relevant quantities present substantial 
time-dependent fluctuations. As may be expected, the general trend of the curves suggests that these fluctuations decrease 
as $\mu$ approaches 1, i.e., the integrable limit. 
We have further ensured that the above trend persists even for different sizes of the lattice. However, in order 
to explore how {\it sensitive} the relevant diagnostics
are towards detecting the breaking of integrability,
we have also performed the computations of
Fig.~\ref{sharp2} which are all conducted in the
vicinity of the integrable limit. Indeed, the
relevant values of $\mu$ are within a range of
less than $5\%$ variations which is a typical limit
where perturbative considerations might be used~\cite{Cai:1994,PhysRevE.68.056603}.
Nevertheless, we can observe that in our extended 
time-horizon evolution dynamics, the relevant quantities
present substantial fluctuations (notice the vertical
axis scale) even very near the integrable limit. 
Indeed, they can be observed even for $\mu=0.99999$
in the case of $C_2$ (although, notably, not in the case
of $C_1$). 
It is only {\it at} the integrable
limit that all relevant such fluctuations disappear
and integrability is retrieved.

An additional aspect that we probe, as the earlier 
results were for $N=100$, is how accurate/sensitive
these diagnostics may be when $N$ is small and the
limited volume of the phase space may not allow to 
probe the potential breaking of integrability. 
We explore this in Fig.~\ref{moment5}
for $N=5$.
This figure, as well as additional tests (not shown) suggest that
it may be easier to ``mistake'' a non-integrable situation
for an integrable one if one uses a very small $N$,
and even more so when one uses a lower moment such as $C_1$.
Already at $N=5$ deviations from integrability are substantially
observable even quite close to $\mu=1$ and $C_2$ turns
out to be a more sensitive probe thereof than $C_1$.

Our conclusion from the above extensive probing of the
parameter space is that these ``former'' conservation
laws constitute a {\it very sensitive} diagnostic
feature of the integrability breaking. 
In fact, higher order such moments (like $C_2$)
are even more sensitive than lower order ones (such as 
$C_1$). Nevertheless, the examination of such quantities,
if such physical/mathematical knowledge is available
can provide a clear measure of deviations from
the ``singular'' (integrable) limit. Nevertheless,
typically such knowledge will, in fact, be absent
(at least, until machine-learning techniques improve
enough to be able to provide such features; see, e.g.,~\cite{PhysRevLett.126.180604} for a recent example).
Thus we are faced with the task of potentially exploring
a more ``generic'' and more widely applicable feature
that could reveal the relevant conservation laws,
their count and eventually the potential integrability
of the nonlinear dynamical system. It is with a view
to the latter direction that we now turn to the (full)
Lyapunov spectrum of the Salerno lattice.

\section{Lyapunov Spectrum and Maximal Lyapunov
Exponent}

The existence of chaotic dynamics is one of the indications of nonintegrability. 
One of the most common tools towards the identification
of such chaoticity consists of the maximal (largest)
Lyapunov exponent (mLCE for short), represented by $\Lambda$,
of the dynamics associated with the
well-known deviation of nearby trajectories.
We compute this prototypical diagnostic by measuring the time evolution of initial perturbations represented by the deviation vector $\chi$~\cite{Mithun:2018,Mithun_Salerno2021} as follows
\begin{equation}
    \Lambda = \lim_{t \rightarrow \infty} L(t),~~\text {where}~~L(t)=\frac{1}{t}\frac{||\chi(t)||}{||\chi(0)||}.
    \label{mlce}
\end{equation}
To obtain a sense of how this diagnostic varies in the
vicinity of integrability, we compute it for a lattice
of $N=100$ particles in Fig.~\ref{mLCE100}. Here, we can
see that the relevant exponent is generically rather
far from the value of $\Lambda \rightarrow 0$, and solely
tends to it in the immediate vicinity of the integrable
limit. In that light, bearing in mind that Hamiltonian
systems feature Lyapunov exponents that are pairwise
symmetric around $0$, if we are aware of such 
Hamiltonian properties of the system, then the vanishing
of this maximal Lyapunov exponent should yield the vanishing
of all of them and hence signal the presence of integrability.

Fig.~\ref{LiapunC1} shows the finite time mLCE $L(t)$ for different values of $N$ and $\mu$. 
{Here, it can be seen that even though
for $N=3$, the system appears to identify $L(t)$ as tending
to $0$ with increasing time, the same is clearly not the case for $N=5$ and beyond, illustrating the non-
integrability of the latter setting.} However, more generally, one may not be aware of the 
Hamiltonian nature of the problem. It may also be
desirable to identify the number of conserved 
quantities of the system. In that vein, we advocate
that a relevant numerical tool consists of the
calculation of the {\it full} Lyapunov spectrum $\lambda_i$
of the system\cite{lichtenberg1992regular}.
A systematic prescription to do so was originally discussed in~\cite{benettin:1980a,benettin:1980b}. Recently,
this was revisited for many-body systems near 
integrability~\cite{merab}. It is relevant to remember in that
context that 
the $\lambda_i$'s corresponding to the conserved quantities are expected to lead to pairs of zeros. 
Hence, we advocate here the usage of the full Lyapunov
spectrum as a generic (i.e., irrespective of the 
details of the system) and straightforward probe of the
number of the system's conservation laws, and ultimately
a sensitive probe of the dynamical lattice's potential
integrability.
Since it is known that the ergodic behavior of the system drastically changes as the number of degrees of the freedom increases \cite{echmann,PhysRevE.104.014218},
we further measure the Lyapunov spectrum as a function of $N$. 

We show the results of the entire Lyapunov spectrum calculation is Fig.~\ref{fig:lspectrum} which is also a central result of
our study. As one expects, the $\lambda_i$'s are zero at $\mu=1$ irrespective of the number of degrees of freedom. 
Naturally, this reflects the integrability of the model
at this parameter value. 
Upon deviation from $\mu=1$, 2 (pairs of) $\lambda_i$'s remain zero due to the two conserved quantities that
are preserved, namely $\mathcal{H}$ and $\mathcal{A}$ as discussed above, 
while the rest become nonzero. It is important to also
note here the role of $N$, the number of degrees of freedom.
When this number is sufficiently low, such as $N=3$, the
limited phase space of the system, in conjunction with the
persisting conservation laws render the recognition
of the non-integrability of the model for $\mu \neq 1$
practically difficult. The same is, more or less, true
for $N=4$, except for parameter values that deviate quite
substantially from $\mu=1$. On the other hand, the example
with $N=5$ and even more demonstrably so the one with $N=10$
make it clear that only two conserved quantities remain
in the evolving dynamics for $\mu \neq 1$ and, hence,
that non-integrability is now prevalent, even though, of course,
these constraints still affect the evolution dynamics.

\section{Conclusions \& Future Challenges}

In the present work we have revisited the Salerno
model as a vehicle for exploring the deviations
from integrability. We have proposed as diagnostics
for monitoring such deviations the examination 
of quantities that are conserved in the integrable
limit, but whose conservation laws are ``broken''
as soon as we depart from that limit. Indeed,
it was found that such quantities are sensitive 
detectors of the deviation from integrability,
incurring large variations even for truly minimal
departures from the Ablowitz-Ladik case, of the
order of $10^{-4}$. This illustrates that relevant
``former conservation laws'' can be used for
monitoring such
integrability-breaking. Nevertheless, we were
subsequently faced with the concern that such
quantities may not be readily available, unless
one has a well-established knowledge of the 
integrable limit. Hence, we sought a set of
quantities that could be classified as generic
and for which computation methods are well-established
that could provide us with a count of the relevant
conservation laws on and off of the integrable limit.
We argued, on the basis of the Salerno example, that the full Lyapunov spectrum is worth considering as a 
a reliable and sufficiently sensitive such
set of quantities, certainly past the limit of very
small degree-of-freedom systems. Indeed, we recalled
that Hamiltonian systems bear Lyapunov exponents
in pairs, and each conservation law leads to a pair
of such exponents that are vanishing, hence the
relevant spectrum is an accurate and (as illustrated
in our case example) sensitive monitor of the number
of conservation laws in a discrete nonlinear dynamical
system such as the Salerno model. Both measurements
of the maximal Lyapunov exponent for different 
numbers of degrees of freedom and levels of proximity
to integrability, and also ones of the full spectrum
were shown to be sensitive to such deviations from
the integrable limit. 

Moving forward, a number of directions for future 
studies are natural to consider. On the one hand,
it seems especially relevant to extend the present
analysis to different systems (discrete and continuum)
to verify the broader relevance of the conclusions 
drawn herein in a larger class of corresponding examples.
Another direction, however, that is equally or even more
promising is that of exploring tools from machine learning
to compute corresponding diagnostics in a fast and efficient
manner. Indeed, in recent years, there has been a substantial
effort towards leveraging such tools to identify underlying
conservation laws~\cite{PhysRevLett.126.180604} and
associated symmetries~\cite{PhysRevLett.128.180201,lamacraft}.
We believe that the diagnostics proposed herein 
(such as the identification of the full Lyapunov
spectrum) are a natural
complement to such efforts and the utilization of such
tools may enable the fast and efficient computation of
such diagnostics even for large(r) number of degree-of-freedom
systems. Such studies are currently in preparation and
will be reported in future publications.

\vspace{3mm} 

{\it Acknowledgements.} This material is based upon work supported by the US
National Science Foundation under Grants No. PHY-2110030 
and DMS-2204702 (P.G.K.), {as well as the Ministry of Education, Science and Technological Development of the Republic of Serbia, grants No.451-03-9/2022-14/ 200017 and 451-03-68/2022-14/200124 (A.Maluckov and A. Man\v ci\'c)}. 
One of us (A.K.) is grateful to Indian
National Science Academy [INSA] for the award of INSA Honarary 
Scientist position at Savitribai Phule Pune University.
The authors acknowledge
insightful discussions with  Alan Bishop, Jesus Cuevas,
Yannis Kevrekidis, Avadh Saxena, Haris Skokos, HongKun Zhang
and Wei Zhu on related topics.

  \bibliographystyle{apsrev4}
\let\itshape\upshape
\normalem
\bibliography{reference1}

\providecommand{\noopsort}[1]{}\providecommand{\singleletter}[1]{#1}%
\begin{thebibliography}{53}%
\makeatletter
\providecommand \@ifxundefined [1]{%
 \@ifx{#1\undefined}
}%
\providecommand \@ifnum [1]{%
 \ifnum #1\expandafter \@firstoftwo
 \else \expandafter \@secondoftwo
 \fi
}%
\providecommand \@ifx [1]{%
 \ifx #1\expandafter \@firstoftwo
 \else \expandafter \@secondoftwo
 \fi
}%
\providecommand \natexlab [1]{#1}%
\providecommand \enquote  [1]{``#1''}%
\providecommand \bibnamefont  [1]{#1}%
\providecommand \bibfnamefont [1]{#1}%
\providecommand \citenamefont [1]{#1}%
\providecommand \href@noop [0]{\@secondoftwo}%
\providecommand \href [0]{\begingroup \@sanitize@url \@href}%
\providecommand \@href[1]{\@@startlink{#1}\@@href}%
\providecommand \@@href[1]{\endgroup#1\@@endlink}%
\providecommand \@sanitize@url [0]{\catcode `\\12\catcode `\$12\catcode
  `\&12\catcode `\#12\catcode `\^12\catcode `\_12\catcode `\%12\relax}%
\providecommand \@@startlink[1]{}%
\providecommand \@@endlink[0]{}%
\providecommand \url  [0]{\begingroup\@sanitize@url \@url }%
\providecommand \@url [1]{\endgroup\@href {#1}{\urlprefix }}%
\providecommand \urlprefix  [0]{URL }%
\providecommand \Eprint [0]{\href }%
\providecommand \doibase [0]{http://dx.doi.org/}%
\providecommand \selectlanguage [0]{\@gobble}%
\providecommand \bibinfo  [0]{\@secondoftwo}%
\providecommand \bibfield  [0]{\@secondoftwo}%
\providecommand \translation [1]{[#1]}%
\providecommand \BibitemOpen [0]{}%
\providecommand \bibitemStop [0]{}%
\providecommand \bibitemNoStop [0]{.\EOS\space}%
\providecommand \EOS [0]{\spacefactor3000\relax}%
\providecommand \BibitemShut  [1]{\csname bibitem#1\endcsname}%
\let\auto@bib@innerbib\@empty
\bibitem [{\citenamefont {Gallavotti}(2008)}]{FPUreview}%
  \BibitemOpen
  \bibfield  {author} {\bibinfo {author} {\bibfnamefont {G.}~\bibnamefont
  {Gallavotti}},\ }\href@noop {} {\emph {\bibinfo {title} {The
  Fermi--Pasta--Ulam Problem: A Status Report}}}\ (\bibinfo  {publisher}
  {Springer-Verlag, Berlin, Germany},\ \bibinfo {year} {2008})\BibitemShut
  {NoStop}%
\bibitem [{\citenamefont {Kevrekidis}(2011)}]{pgk:2011}%
  \BibitemOpen
  \bibfield  {author} {\bibinfo {author} {\bibfnamefont {P.}~\bibnamefont
  {Kevrekidis}},\ }\bibfield  {title} {\enquote {\bibinfo {title} {Non-linear
  waves in lattices: past, present, future},}\ }\href@noop {} {\bibfield
  {journal} {\bibinfo  {journal} {IMA J. Appl. Math.}\ }\textbf {\bibinfo
  {volume} {76}},\ \bibinfo {pages} {389} (\bibinfo {year} {2011})}\BibitemShut
  {NoStop}%
\bibitem [{\citenamefont {Sievers}\ and\ \citenamefont {Takeno}(1988)}]{ST}%
  \BibitemOpen
  \bibfield  {author} {\bibinfo {author} {\bibfnamefont {A.~J.}\ \bibnamefont
  {Sievers}}\ and\ \bibinfo {author} {\bibfnamefont {S.}~\bibnamefont
  {Takeno}},\ }\bibfield  {title} {\enquote {\bibinfo {title} {Intrinsic
  localized modes in anharmonic crystals},}\ }\href@noop {} {\bibfield
  {journal} {\bibinfo  {journal} {Phys. Rev. Lett.}\ }\textbf {\bibinfo
  {volume} {61}},\ \bibinfo {pages} {970} (\bibinfo {year} {1988})}\BibitemShut
  {NoStop}%
\bibitem [{\citenamefont {Page}(1990)}]{Page}%
  \BibitemOpen
  \bibfield  {author} {\bibinfo {author} {\bibfnamefont {J.~B.}\ \bibnamefont
  {Page}},\ }\bibfield  {title} {\enquote {\bibinfo {title} {Asymptotic
  solutions for localized vibrational modes in strongly anharmonic periodic
  systems},}\ }\href {\doibase 10.1103/PhysRevB.41.7835} {\bibfield  {journal}
  {\bibinfo  {journal} {Phys. Rev. B}\ }\textbf {\bibinfo {volume} {41}},\
  \bibinfo {pages} {7835} (\bibinfo {year} {1990})}\BibitemShut {NoStop}%
\bibitem [{\citenamefont {Flach}\ and\ \citenamefont
  {Gorbach}(2008)}]{Flach:2008}%
  \BibitemOpen
  \bibfield  {author} {\bibinfo {author} {\bibfnamefont {S.}~\bibnamefont
  {Flach}}\ and\ \bibinfo {author} {\bibfnamefont {A.~V.}\ \bibnamefont
  {Gorbach}},\ }\bibfield  {title} {\enquote {\bibinfo {title} {Discrete
  breathers - {A}dvances in theory and applications},}\ }\href {\doibase
  10.1016/j.physrep.2008.05.002} {\bibfield  {journal} {\bibinfo  {journal}
  {Phys. Rep.}\ }\textbf {\bibinfo {volume} {467}},\ \bibinfo {pages} {1 }
  (\bibinfo {year} {2008})}\BibitemShut {NoStop}%
\bibitem [{\citenamefont {Lederer}\ \emph {et~al.}(2008)\citenamefont
  {Lederer}, \citenamefont {Stegeman}, \citenamefont {Christodoulides},
  \citenamefont {Assanto}, \citenamefont {Segev},\ and\ \citenamefont
  {Silberberg}}]{moti}%
  \BibitemOpen
  \bibfield  {author} {\bibinfo {author} {\bibfnamefont {F.}~\bibnamefont
  {Lederer}}, \bibinfo {author} {\bibfnamefont {G.~I.}\ \bibnamefont
  {Stegeman}}, \bibinfo {author} {\bibfnamefont {D.~N.}\ \bibnamefont
  {Christodoulides}}, \bibinfo {author} {\bibfnamefont {G.}~\bibnamefont
  {Assanto}}, \bibinfo {author} {\bibfnamefont {M.}~\bibnamefont {Segev}}, \
  and\ \bibinfo {author} {\bibfnamefont {Y.}~\bibnamefont {Silberberg}},\
  }\bibfield  {title} {\enquote {\bibinfo {title} {Discrete solitons in
  optics},}\ }\href@noop {} {\bibfield  {journal} {\bibinfo  {journal} {Phys.
  Rep.}\ }\textbf {\bibinfo {volume} {463}},\ \bibinfo {pages} {1} (\bibinfo
  {year} {2008})}\BibitemShut {NoStop}%
\bibitem [{\citenamefont {Morsch}\ and\ \citenamefont
  {Oberthaler}(2006)}]{Morsch}%
  \BibitemOpen
  \bibfield  {author} {\bibinfo {author} {\bibfnamefont {O.}~\bibnamefont
  {Morsch}}\ and\ \bibinfo {author} {\bibfnamefont {M.}~\bibnamefont
  {Oberthaler}},\ }\bibfield  {title} {\enquote {\bibinfo {title} {Dynamics of
  {B}ose--{E}instein condensates in optical lattices},}\ }\href@noop {}
  {\bibfield  {journal} {\bibinfo  {journal} {Rev. Mod. Phys.}\ }\textbf
  {\bibinfo {volume} {78}},\ \bibinfo {pages} {179} (\bibinfo {year}
  {2006})}\BibitemShut {NoStop}%
\bibitem [{\citenamefont {Sato}\ \emph {et~al.}(2006)\citenamefont {Sato},
  \citenamefont {Hubbard},\ and\ \citenamefont {Sievers}}]{sievers}%
  \BibitemOpen
  \bibfield  {author} {\bibinfo {author} {\bibfnamefont {M.}~\bibnamefont
  {Sato}}, \bibinfo {author} {\bibfnamefont {B.~E.}\ \bibnamefont {Hubbard}}, \
  and\ \bibinfo {author} {\bibfnamefont {A.~J.}\ \bibnamefont {Sievers}},\
  }\bibfield  {title} {\enquote {\bibinfo {title} {\textit{Colloquium}:
  Nonlinear energy localization and its manipulation in micromechanical
  oscillator arrays},}\ }\href@noop {} {\bibfield  {journal} {\bibinfo
  {journal} {Rev. Mod. Phys.}\ }\textbf {\bibinfo {volume} {78}},\ \bibinfo
  {pages} {137} (\bibinfo {year} {2006})}\BibitemShut {NoStop}%
\bibitem [{\citenamefont {Starosvetsky}\ \emph {et~al.}(2017)\citenamefont
  {Starosvetsky}, \citenamefont {Jayaprakash}, \citenamefont {Hasan},\ and\
  \citenamefont {Vakakis}}]{yuli_book}%
  \BibitemOpen
  \bibfield  {author} {\bibinfo {author} {\bibfnamefont {Y.}~\bibnamefont
  {Starosvetsky}}, \bibinfo {author} {\bibfnamefont {K.}~\bibnamefont
  {Jayaprakash}}, \bibinfo {author} {\bibfnamefont {M.~A.}\ \bibnamefont
  {Hasan}}, \ and\ \bibinfo {author} {\bibfnamefont {A.}~\bibnamefont
  {Vakakis}},\ }\href@noop {} {\emph {\bibinfo {title} {Dynamics and Acoustics
  of Ordered Granular Media}}}\ (\bibinfo  {publisher} {World Scientific,
  Singapore},\ \bibinfo {year} {2017})\BibitemShut {NoStop}%
\bibitem [{\citenamefont {Chong}\ and\ \citenamefont
  {Kevrekidis}(2018)}]{granularBook}%
  \BibitemOpen
  \bibfield  {author} {\bibinfo {author} {\bibfnamefont {C.}~\bibnamefont
  {Chong}}\ and\ \bibinfo {author} {\bibfnamefont {P.~G.}\ \bibnamefont
  {Kevrekidis}},\ }\href@noop {} {\emph {\bibinfo {title} {Coherent Structures
  in Granular Crystals: From Experiment and Modelling to Computation and
  Mathematical Analysis}}}\ (\bibinfo  {publisher} {Springer},\ \bibinfo
  {address} {New York},\ \bibinfo {year} {2018})\BibitemShut {NoStop}%
\bibitem [{\citenamefont {Remoissenet}(1999)}]{remoissenet}%
  \BibitemOpen
  \bibfield  {author} {\bibinfo {author} {\bibfnamefont {M.}~\bibnamefont
  {Remoissenet}},\ }\href@noop {} {\emph {\bibinfo {title} {Waves Called
  Solitons}}}\ (\bibinfo  {publisher} {Springer-Verlag, Berlin},\ \bibinfo
  {year} {1999})\BibitemShut {NoStop}%
\bibitem [{\citenamefont {English}\ \emph {et~al.}(2003)\citenamefont
  {English}, \citenamefont {Sato},\ and\ \citenamefont {Sievers}}]{lars3}%
  \BibitemOpen
  \bibfield  {author} {\bibinfo {author} {\bibfnamefont {L.~Q.}\ \bibnamefont
  {English}}, \bibinfo {author} {\bibfnamefont {M.}~\bibnamefont {Sato}}, \
  and\ \bibinfo {author} {\bibfnamefont {A.~J.}\ \bibnamefont {Sievers}},\
  }\bibfield  {title} {\enquote {\bibinfo {title} {Modulational instability of
  nonlinear spin waves in easy-axis antiferromagnetic chains. ii. influence of
  sample shape on intrinsic localized modes and dynamic spin defects},}\
  }\href@noop {} {\bibfield  {journal} {\bibinfo  {journal} {Phys. Rev. B}\
  }\textbf {\bibinfo {volume} {67}},\ \bibinfo {pages} {024403} (\bibinfo
  {year} {2003})}\BibitemShut {NoStop}%
\bibitem [{\citenamefont {Schwarz}\ \emph {et~al.}(1999)\citenamefont
  {Schwarz}, \citenamefont {English},\ and\ \citenamefont {Sievers}}]{lars4}%
  \BibitemOpen
  \bibfield  {author} {\bibinfo {author} {\bibfnamefont {U.~T.}\ \bibnamefont
  {Schwarz}}, \bibinfo {author} {\bibfnamefont {L.~Q.}\ \bibnamefont
  {English}}, \ and\ \bibinfo {author} {\bibfnamefont {A.~J.}\ \bibnamefont
  {Sievers}},\ }\bibfield  {title} {\enquote {\bibinfo {title} {Experimental
  generation and observation of intrinsic localized spin wave modes in an
  antiferromagnet},}\ }\href@noop {} {\bibfield  {journal} {\bibinfo  {journal}
  {Phys. Rev. Lett.}\ }\textbf {\bibinfo {volume} {83}},\ \bibinfo {pages}
  {223} (\bibinfo {year} {1999})}\BibitemShut {NoStop}%
\bibitem [{\citenamefont {Binder}\ \emph {et~al.}(2000)\citenamefont {Binder},
  \citenamefont {Abraimov}, \citenamefont {Ustinov}, \citenamefont {Flach},\
  and\ \citenamefont {Zolotaryuk}}]{alex}%
  \BibitemOpen
  \bibfield  {author} {\bibinfo {author} {\bibfnamefont {P.}~\bibnamefont
  {Binder}}, \bibinfo {author} {\bibfnamefont {D.}~\bibnamefont {Abraimov}},
  \bibinfo {author} {\bibfnamefont {A.~V.}\ \bibnamefont {Ustinov}}, \bibinfo
  {author} {\bibfnamefont {S.}~\bibnamefont {Flach}}, \ and\ \bibinfo {author}
  {\bibfnamefont {Y.}~\bibnamefont {Zolotaryuk}},\ }\bibfield  {title}
  {\enquote {\bibinfo {title} {Observation of breathers in {Josephson}
  ladders},}\ }\href@noop {} {\bibfield  {journal} {\bibinfo  {journal} {Phys.
  Rev. Lett.}\ }\textbf {\bibinfo {volume} {84}},\ \bibinfo {pages} {745}
  (\bibinfo {year} {2000})}\BibitemShut {NoStop}%
\bibitem [{\citenamefont {Tr\'{\i}as}\ \emph {et~al.}(2000)\citenamefont
  {Tr\'{\i}as}, \citenamefont {Mazo},\ and\ \citenamefont {Orlando}}]{alex2}%
  \BibitemOpen
  \bibfield  {author} {\bibinfo {author} {\bibfnamefont {E.}~\bibnamefont
  {Tr\'{\i}as}}, \bibinfo {author} {\bibfnamefont {J.~J.}\ \bibnamefont
  {Mazo}}, \ and\ \bibinfo {author} {\bibfnamefont {T.~P.}\ \bibnamefont
  {Orlando}},\ }\bibfield  {title} {\enquote {\bibinfo {title} {Discrete
  breathers in nonlinear lattices: Experimental detection in a josephson
  array},}\ }\href@noop {} {\bibfield  {journal} {\bibinfo  {journal} {Phys.
  Rev. Lett.}\ }\textbf {\bibinfo {volume} {84}},\ \bibinfo {pages} {741}
  (\bibinfo {year} {2000})}\BibitemShut {NoStop}%
\bibitem [{\citenamefont {Peyrard}(2004)}]{Peybi}%
  \BibitemOpen
  \bibfield  {author} {\bibinfo {author} {\bibfnamefont {M.}~\bibnamefont
  {Peyrard}},\ }\bibfield  {title} {\enquote {\bibinfo {title} {Nonlinear
  dynamics and statistical physics of {DNA}},}\ }\href@noop {} {\bibfield
  {journal} {\bibinfo  {journal} {Nonlinearity}\ }\textbf {\bibinfo {volume}
  {17}},\ \bibinfo {pages} {R1} (\bibinfo {year} {2004})}\BibitemShut {NoStop}%
\bibitem [{\citenamefont {Aubry}(2006)}]{Aubry06}%
  \BibitemOpen
  \bibfield  {author} {\bibinfo {author} {\bibfnamefont {S.}~\bibnamefont
  {Aubry}},\ }\bibfield  {title} {\enquote {\bibinfo {title} {Discrete
  breathers: {L}ocalization and transfer of energy in discrete {H}amiltonian
  nonlinear systems},}\ }\href@noop {} {\bibfield  {journal} {\bibinfo
  {journal} {Physica D}\ }\textbf {\bibinfo {volume} {216}},\ \bibinfo {pages}
  {1} (\bibinfo {year} {2006})}\BibitemShut {NoStop}%
\bibitem [{\citenamefont {Kevrekidis}(2009)}]{kev2009}%
  \BibitemOpen
  \bibfield  {author} {\bibinfo {author} {\bibfnamefont {P.}~\bibnamefont
  {Kevrekidis}},\ }\href@noop {} {\emph {\bibinfo {title} {The discrete
  nonlinear Schr{\"o}dinger equation}}},\ Vol.\ \bibinfo {volume} {232}\
  (\bibinfo  {publisher} {Springer Science \& Business Media},\ \bibinfo {year}
  {2009})\BibitemShut {NoStop}%
\bibitem [{\citenamefont {Fermi}\ \emph {et~al.}(1955)\citenamefont {Fermi},
  \citenamefont {Pasta},\ and\ \citenamefont {Ulam}}]{Fermi:1955}%
  \BibitemOpen
  \bibfield  {author} {\bibinfo {author} {\bibfnamefont {E.}~\bibnamefont
  {Fermi}}, \bibinfo {author} {\bibfnamefont {J.}~\bibnamefont {Pasta}}, \ and\
  \bibinfo {author} {\bibfnamefont {S.}~\bibnamefont {Ulam}},\ }\bibfield
  {title} {\enquote {\bibinfo {title} {Los {A}lamos {S}cientific {L}aboratory
  {R}eport},}\ }\href@noop {} {\bibfield  {journal} {\bibinfo  {journal}
  {LA-1940}\ } (\bibinfo {year} {1955})}\BibitemShut {NoStop}%
\bibitem [{\citenamefont {Danieli}\ \emph {et~al.}(2017)\citenamefont
  {Danieli}, \citenamefont {Campbell},\ and\ \citenamefont
  {Flach}}]{Danieli:2017}%
  \BibitemOpen
  \bibfield  {author} {\bibinfo {author} {\bibfnamefont {C.}~\bibnamefont
  {Danieli}}, \bibinfo {author} {\bibfnamefont {D.~K.}\ \bibnamefont
  {Campbell}}, \ and\ \bibinfo {author} {\bibfnamefont {S.}~\bibnamefont
  {Flach}},\ }\bibfield  {title} {\enquote {\bibinfo {title} {Intermittent
  many-body dynamics at equilibrium},}\ }\href {\doibase
  10.1103/PhysRevE.95.060202} {\bibfield  {journal} {\bibinfo  {journal} {Phys.
  Rev. E}\ }\textbf {\bibinfo {volume} {95}},\ \bibinfo {pages} {060202}
  (\bibinfo {year} {2017})}\BibitemShut {NoStop}%
\bibitem [{\citenamefont {Danieli}\ \emph
  {et~al.}(2019{\natexlab{a}})\citenamefont {Danieli}, \citenamefont {Mithun},
  \citenamefont {Kati}, \citenamefont {Campbell},\ and\ \citenamefont
  {Flach}}]{Danieli:2019}%
  \BibitemOpen
  \bibfield  {author} {\bibinfo {author} {\bibfnamefont {C.}~\bibnamefont
  {Danieli}}, \bibinfo {author} {\bibfnamefont {T.}~\bibnamefont {Mithun}},
  \bibinfo {author} {\bibfnamefont {Y.}~\bibnamefont {Kati}}, \bibinfo {author}
  {\bibfnamefont {D.~K.}\ \bibnamefont {Campbell}}, \ and\ \bibinfo {author}
  {\bibfnamefont {S.}~\bibnamefont {Flach}},\ }\bibfield  {title} {\enquote
  {\bibinfo {title} {Dynamical glass in weakly nonintegrable klein-gordon
  chains},}\ }\href {\doibase 10.1103/PhysRevE.100.032217} {\bibfield
  {journal} {\bibinfo  {journal} {Phys. Rev. E}\ }\textbf {\bibinfo {volume}
  {100}},\ \bibinfo {pages} {032217} (\bibinfo {year}
  {2019}{\natexlab{a}})}\BibitemShut {NoStop}%
\bibitem [{\citenamefont {Mithun}\ \emph {et~al.}(2019)\citenamefont {Mithun},
  \citenamefont {Danieli}, \citenamefont {Kati},\ and\ \citenamefont
  {Flach}}]{Mithun:2019}%
  \BibitemOpen
  \bibfield  {author} {\bibinfo {author} {\bibfnamefont {T.}~\bibnamefont
  {Mithun}}, \bibinfo {author} {\bibfnamefont {C.}~\bibnamefont {Danieli}},
  \bibinfo {author} {\bibfnamefont {Y.}~\bibnamefont {Kati}}, \ and\ \bibinfo
  {author} {\bibfnamefont {S.}~\bibnamefont {Flach}},\ }\bibfield  {title}
  {\enquote {\bibinfo {title} {Dynamical glass and ergodization times in
  classical josephson junction chains},}\ }\href {\doibase
  10.1103/PhysRevLett.122.054102} {\bibfield  {journal} {\bibinfo  {journal}
  {Phys. Rev. Lett.}\ }\textbf {\bibinfo {volume} {122}},\ \bibinfo {pages}
  {054102} (\bibinfo {year} {2019})}\BibitemShut {NoStop}%
\bibitem [{\citenamefont {Malishava}\ and\ \citenamefont
  {Flach}(2022)}]{merab}%
  \BibitemOpen
  \bibfield  {author} {\bibinfo {author} {\bibfnamefont {M.}~\bibnamefont
  {Malishava}}\ and\ \bibinfo {author} {\bibfnamefont {S.}~\bibnamefont
  {Flach}},\ }\bibfield  {title} {\enquote {\bibinfo {title} {Lyapunov spectrum
  scaling for classical many-body dynamics close to integrability},}\ }\href
  {\doibase 10.1103/PhysRevLett.128.134102} {\bibfield  {journal} {\bibinfo
  {journal} {Phys. Rev. Lett.}\ }\textbf {\bibinfo {volume} {128}},\ \bibinfo
  {pages} {134102} (\bibinfo {year} {2022})}\BibitemShut {NoStop}%
\bibitem [{\citenamefont {Zabusky}\ and\ \citenamefont
  {Kruskal}(1965)}]{Zabusky:1965}%
  \BibitemOpen
  \bibfield  {author} {\bibinfo {author} {\bibfnamefont {N.~J.}\ \bibnamefont
  {Zabusky}}\ and\ \bibinfo {author} {\bibfnamefont {M.~D.}\ \bibnamefont
  {Kruskal}},\ }\bibfield  {title} {\enquote {\bibinfo {title} {Interaction of
  "solitons" in a collisionless plasma and the recurrence of initial states},}\
  }\href {https://journals.aps.org/prl/pdf/10.1103/PhysRevLett.15.240}
  {\bibfield  {journal} {\bibinfo  {journal} {Phys. Rev. Lett.}\ }\textbf
  {\bibinfo {volume} {15}},\ \bibinfo {pages} {240} (\bibinfo {year}
  {1965})}\BibitemShut {NoStop}%
\bibitem [{\citenamefont {Porter}\ \emph {et~al.}(2009)\citenamefont {Porter},
  \citenamefont {Zabusky}, \citenamefont {Hu},\ and\ \citenamefont
  {Campbell}}]{Porter:2009}%
  \BibitemOpen
  \bibfield  {author} {\bibinfo {author} {\bibfnamefont {M.~A.}\ \bibnamefont
  {Porter}}, \bibinfo {author} {\bibfnamefont {N.~J.}\ \bibnamefont {Zabusky}},
  \bibinfo {author} {\bibfnamefont {B.}~\bibnamefont {Hu}}, \ and\ \bibinfo
  {author} {\bibfnamefont {D.~K.}\ \bibnamefont {Campbell}},\ }\bibfield
  {title} {\enquote {\bibinfo {title} {Fermi, {P}asta, {U}lam and the birth of
  experimental mathematics},}\ }\href
  {https://people.maths.ox.ac.uk/porterm/papers/fpupop_final.pdf} {\bibfield
  {journal} {\bibinfo  {journal} {Am. Sci.}\ }\textbf {\bibinfo {volume} {97}}
  (\bibinfo {year} {2009})}\BibitemShut {NoStop}%
\bibitem [{\citenamefont {Ablowitz}\ and\ \citenamefont {Segur}(1981)}]{AS81}%
  \BibitemOpen
  \bibfield  {author} {\bibinfo {author} {\bibfnamefont {M.}~\bibnamefont
  {Ablowitz}}\ and\ \bibinfo {author} {\bibfnamefont {H.}~\bibnamefont
  {Segur}},\ }\href@noop {} {\emph {\bibinfo {title} {Solitons and the inverse
  scattering transform}}},\ Vol.~\bibinfo {volume} {4}\ (\bibinfo  {publisher}
  {SIAM},\ \bibinfo {address} {Philadelphia},\ \bibinfo {year}
  {1981})\BibitemShut {NoStop}%
\bibitem [{\citenamefont {Ablowitz}(2011)}]{ablowitz2}%
  \BibitemOpen
  \bibfield  {author} {\bibinfo {author} {\bibfnamefont {M.}~\bibnamefont
  {Ablowitz}},\ }\href@noop {} {\emph {\bibinfo {title} {Nonlinear Dispersive
  Waves, Asymptotic Analysis and Solitons}}}\ (\bibinfo  {publisher} {Cambridge
  University Press, Cambridge},\ \bibinfo {year} {2011})\BibitemShut {NoStop}%
\bibitem [{\citenamefont {Kivshar}\ and\ \citenamefont
  {Malomed}(1989)}]{RevModPhys.61.763}%
  \BibitemOpen
  \bibfield  {author} {\bibinfo {author} {\bibfnamefont {Y.~S.}\ \bibnamefont
  {Kivshar}}\ and\ \bibinfo {author} {\bibfnamefont {B.~A.}\ \bibnamefont
  {Malomed}},\ }\bibfield  {title} {\enquote {\bibinfo {title} {Dynamics of
  solitons in nearly integrable systems},}\ }\href {\doibase
  10.1103/RevModPhys.61.763} {\bibfield  {journal} {\bibinfo  {journal} {Rev.
  Mod. Phys.}\ }\textbf {\bibinfo {volume} {61}},\ \bibinfo {pages} {763}
  (\bibinfo {year} {1989})}\BibitemShut {NoStop}%
\bibitem [{\citenamefont {Hennig}\ \emph
  {et~al.}(2022{\natexlab{a}})\citenamefont {Hennig}, \citenamefont
  {Karachalios},\ and\ \citenamefont {Cuevas-Maraver}}]{karch1}%
  \BibitemOpen
  \bibfield  {author} {\bibinfo {author} {\bibfnamefont {D.}~\bibnamefont
  {Hennig}}, \bibinfo {author} {\bibfnamefont {N.~I.}\ \bibnamefont
  {Karachalios}}, \ and\ \bibinfo {author} {\bibfnamefont {J.}~\bibnamefont
  {Cuevas-Maraver}},\ }\bibfield  {title} {\enquote {\bibinfo {title} {The
  closeness of localized structures between the ablowitz-ladik lattice and
  discrete nonlinear schr{\"o}dinger equations: Generalized al and dnls
  systems},}\ }\href {\doibase 10.1063/5.0072391} {\bibfield  {journal}
  {\bibinfo  {journal} {Journal of Mathematical Physics}\ }\textbf {\bibinfo
  {volume} {63}},\ \bibinfo {pages} {042701} (\bibinfo {year}
  {2022}{\natexlab{a}})},\ \Eprint
  {http://arxiv.org/abs/https://doi.org/10.1063/5.0072391}
  {https://doi.org/10.1063/5.0072391} \BibitemShut {NoStop}%
\bibitem [{\citenamefont {Hennig}\ \emph
  {et~al.}(2022{\natexlab{b}})\citenamefont {Hennig}, \citenamefont
  {Karachalios},\ and\ \citenamefont {Cuevas-Maraver}}]{karch2}%
  \BibitemOpen
  \bibfield  {author} {\bibinfo {author} {\bibfnamefont {D.}~\bibnamefont
  {Hennig}}, \bibinfo {author} {\bibfnamefont {N.~I.}\ \bibnamefont
  {Karachalios}}, \ and\ \bibinfo {author} {\bibfnamefont {J.}~\bibnamefont
  {Cuevas-Maraver}},\ }\bibfield  {title} {\enquote {\bibinfo {title} {The
  closeness of the ablowitz-ladik lattice to the discrete nonlinear
  schr{\"o}dinger equation},}\ }\href {\doibase
  https://doi.org/10.1016/j.jde.2022.01.050} {\bibfield  {journal} {\bibinfo
  {journal} {Journal of Differential Equations}\ }\textbf {\bibinfo {volume}
  {316}},\ \bibinfo {pages} {346} (\bibinfo {year}
  {2022}{\natexlab{b}})}\BibitemShut {NoStop}%
\bibitem [{\citenamefont {Dmitriev}\ \emph {et~al.}(2003)\citenamefont
  {Dmitriev}, \citenamefont {Kevrekidis}, \citenamefont {Malomed},\ and\
  \citenamefont {Frantzeskakis}}]{PhysRevE.68.056603}%
  \BibitemOpen
  \bibfield  {author} {\bibinfo {author} {\bibfnamefont {S.~V.}\ \bibnamefont
  {Dmitriev}}, \bibinfo {author} {\bibfnamefont {P.~G.}\ \bibnamefont
  {Kevrekidis}}, \bibinfo {author} {\bibfnamefont {B.~A.}\ \bibnamefont
  {Malomed}}, \ and\ \bibinfo {author} {\bibfnamefont {D.~J.}\ \bibnamefont
  {Frantzeskakis}},\ }\bibfield  {title} {\enquote {\bibinfo {title}
  {Two-soliton collisions in a near-integrable lattice system},}\ }\href
  {\doibase 10.1103/PhysRevE.68.056603} {\bibfield  {journal} {\bibinfo
  {journal} {Phys. Rev. E}\ }\textbf {\bibinfo {volume} {68}},\ \bibinfo
  {pages} {056603} (\bibinfo {year} {2003})}\BibitemShut {NoStop}%
\bibitem [{\citenamefont {Salerno}(1992)}]{salerno1992quantum}%
  \BibitemOpen
  \bibfield  {author} {\bibinfo {author} {\bibfnamefont {M.}~\bibnamefont
  {Salerno}},\ }\bibfield  {title} {\enquote {\bibinfo {title} {Quantum
  deformations of the discrete nonlinear schr{\"o}dinger equation},}\
  }\href@noop {} {\bibfield  {journal} {\bibinfo  {journal} {Physical Review
  A}\ }\textbf {\bibinfo {volume} {46}},\ \bibinfo {pages} {6856} (\bibinfo
  {year} {1992})}\BibitemShut {NoStop}%
\bibitem [{\citenamefont {Ablowitz}\ and\ \citenamefont
  {Ladik}(1976)}]{ablowitz1976nonlinear}%
  \BibitemOpen
  \bibfield  {author} {\bibinfo {author} {\bibfnamefont {M.}~\bibnamefont
  {Ablowitz}}\ and\ \bibinfo {author} {\bibfnamefont {J.}~\bibnamefont
  {Ladik}},\ }\bibfield  {title} {\enquote {\bibinfo {title} {Nonlinear
  differential--difference equations and fourier analysis},}\ }\href@noop {}
  {\bibfield  {journal} {\bibinfo  {journal} {Journal of Mathematical Physics}\
  }\textbf {\bibinfo {volume} {17}},\ \bibinfo {pages} {1011} (\bibinfo {year}
  {1976})}\BibitemShut {NoStop}%
\bibitem [{\citenamefont {Ablowitz}\ \emph {et~al.}(2003)\citenamefont
  {Ablowitz}, \citenamefont {Prinari},\ and\ \citenamefont
  {Trubatch}}]{ablowitz_prinari_trubatch_2003}%
  \BibitemOpen
  \bibfield  {author} {\bibinfo {author} {\bibfnamefont {M.~J.}\ \bibnamefont
  {Ablowitz}}, \bibinfo {author} {\bibfnamefont {B.}~\bibnamefont {Prinari}}, \
  and\ \bibinfo {author} {\bibfnamefont {A.~D.}\ \bibnamefont {Trubatch}},\
  }\href {\doibase 10.1017/CBO9780511546709} {\emph {\bibinfo {title} {Discrete
  and Continuous Nonlinear Schr\"odinger Systems}}},\ London Mathematical
  Society Lecture Note Series\ (\bibinfo  {publisher} {Cambridge University
  Press},\ \bibinfo {year} {2003})\BibitemShut {NoStop}%
\bibitem [{\citenamefont {Benettin}\ \emph
  {et~al.}(1980{\natexlab{a}})\citenamefont {Benettin}, \citenamefont
  {Galgani}, \citenamefont {Giorgilli},\ and\ \citenamefont
  {Strelcyn}}]{benettin:1980a}%
  \BibitemOpen
  \bibfield  {author} {\bibinfo {author} {\bibfnamefont {G.}~\bibnamefont
  {Benettin}}, \bibinfo {author} {\bibfnamefont {L.}~\bibnamefont {Galgani}},
  \bibinfo {author} {\bibfnamefont {A.}~\bibnamefont {Giorgilli}}, \ and\
  \bibinfo {author} {\bibfnamefont {J.-M.}\ \bibnamefont {Strelcyn}},\
  }\bibfield  {title} {\enquote {\bibinfo {title} {Lyapunov characteristic
  exponents for smooth dynamical systems and for {H}amiltonian systems; a
  method for computing all of them. {P}art 1: {T}heory},}\ }\href {\doibase
  10.1007/BF02128236} {\bibfield  {journal} {\bibinfo  {journal} {Meccanica}\
  }\textbf {\bibinfo {volume} {15}},\ \bibinfo {pages} {9} (\bibinfo {year}
  {1980}{\natexlab{a}})}\BibitemShut {NoStop}%
\bibitem [{\citenamefont {Benettin}\ \emph
  {et~al.}(1980{\natexlab{b}})\citenamefont {Benettin}, \citenamefont
  {Galgani}, \citenamefont {Giorgilli},\ and\ \citenamefont
  {Strelcyn}}]{benettin:1980b}%
  \BibitemOpen
  \bibfield  {author} {\bibinfo {author} {\bibfnamefont {G.}~\bibnamefont
  {Benettin}}, \bibinfo {author} {\bibfnamefont {L.}~\bibnamefont {Galgani}},
  \bibinfo {author} {\bibfnamefont {A.}~\bibnamefont {Giorgilli}}, \ and\
  \bibinfo {author} {\bibfnamefont {J.}~\bibnamefont {Strelcyn}},\ }\bibfield
  {title} {\enquote {\bibinfo {title} {Lyapunov characteristic exponents for
  smooth dynamical systems; a method for computing all of them. {P}art 2:
  {N}umerical application},}\ }\href {\doibase 10.1007/BF02128237} {\bibfield
  {journal} {\bibinfo  {journal} {Meccanica}\ }\textbf {\bibinfo {volume}
  {15}},\ \bibinfo {pages} {21} (\bibinfo {year}
  {1980}{\natexlab{b}})}\BibitemShut {NoStop}%
\bibitem [{\citenamefont {Cai}\ \emph {et~al.}(1996)\citenamefont {Cai},
  \citenamefont {Bishop},\ and\ \citenamefont
  {Gr\o{}nbech-Jensen}}]{PhysRevE.53.4131}%
  \BibitemOpen
  \bibfield  {author} {\bibinfo {author} {\bibfnamefont {D.}~\bibnamefont
  {Cai}}, \bibinfo {author} {\bibfnamefont {A.~R.}\ \bibnamefont {Bishop}}, \
  and\ \bibinfo {author} {\bibfnamefont {N.}~\bibnamefont
  {Gr\o{}nbech-Jensen}},\ }\bibfield  {title} {\enquote {\bibinfo {title}
  {Perturbation theories of a discrete, integrable nonlinear schr\"odinger
  equation},}\ }\href {\doibase 10.1103/PhysRevE.53.4131} {\bibfield  {journal}
  {\bibinfo  {journal} {Phys. Rev. E}\ }\textbf {\bibinfo {volume} {53}},\
  \bibinfo {pages} {4131} (\bibinfo {year} {1996})}\BibitemShut {NoStop}%
\bibitem [{\citenamefont {Gomez-Gardenes}\ \emph {et~al.}(2004)\citenamefont
  {Gomez-Gardenes}, \citenamefont {Falo},\ and\ \citenamefont
  {Floria}}]{GOMEZGARDENES2004213}%
  \BibitemOpen
  \bibfield  {author} {\bibinfo {author} {\bibfnamefont {J.}~\bibnamefont
  {Gomez-Gardenes}}, \bibinfo {author} {\bibfnamefont {F.}~\bibnamefont
  {Falo}}, \ and\ \bibinfo {author} {\bibfnamefont {L.}~\bibnamefont
  {Floria}},\ }\bibfield  {title} {\enquote {\bibinfo {title} {Mobile
  localization in nonlinear schr\"odinger lattices},}\ }\href {\doibase
  https://doi.org/10.1016/j.physleta.2004.09.049} {\bibfield  {journal}
  {\bibinfo  {journal} {Physics Letters A}\ }\textbf {\bibinfo {volume}
  {332}},\ \bibinfo {pages} {213} (\bibinfo {year} {2004})}\BibitemShut
  {NoStop}%
\bibitem [{\citenamefont {Mithun}\ \emph
  {et~al.}(2021{\natexlab{a}})\citenamefont {Mithun}, \citenamefont {Maluckov},
  \citenamefont {Manda}, \citenamefont {Skokos}, \citenamefont {Bishop},
  \citenamefont {Saxena}, \citenamefont {Khare},\ and\ \citenamefont
  {Kevrekidis}}]{Mithun_Salerno2021}%
  \BibitemOpen
  \bibfield  {author} {\bibinfo {author} {\bibfnamefont {T.}~\bibnamefont
  {Mithun}}, \bibinfo {author} {\bibfnamefont {A.}~\bibnamefont {Maluckov}},
  \bibinfo {author} {\bibfnamefont {B.~M.}\ \bibnamefont {Manda}}, \bibinfo
  {author} {\bibfnamefont {C.}~\bibnamefont {Skokos}}, \bibinfo {author}
  {\bibfnamefont {A.}~\bibnamefont {Bishop}}, \bibinfo {author} {\bibfnamefont
  {A.}~\bibnamefont {Saxena}}, \bibinfo {author} {\bibfnamefont
  {A.}~\bibnamefont {Khare}}, \ and\ \bibinfo {author} {\bibfnamefont {P.~G.}\
  \bibnamefont {Kevrekidis}},\ }\bibfield  {title} {\enquote {\bibinfo {title}
  {Thermalization in the one-dimensional salerno model lattice},}\ }\href
  {\doibase 10.1103/PhysRevE.103.032211} {\bibfield  {journal} {\bibinfo
  {journal} {Phys. Rev. E}\ }\textbf {\bibinfo {volume} {103}},\ \bibinfo
  {pages} {032211} (\bibinfo {year} {2021}{\natexlab{a}})}\BibitemShut
  {NoStop}%
\bibitem [{\citenamefont {Cassidy}(2010)}]{cassidy2010chaos}%
  \BibitemOpen
  \bibfield  {author} {\bibinfo {author} {\bibfnamefont {A.~C.}\ \bibnamefont
  {Cassidy}},\ }\bibfield  {title} {\enquote {\bibinfo {title} {Chaos and
  thermalization in the one-dimensional bose-hubbard model in the
  classical-field approximation},}\ }\href@noop {} {\bibfield  {journal}
  {\bibinfo  {journal} {arXiv preprint arXiv:1003.5206}\ } (\bibinfo {year}
  {2010})}\BibitemShut {NoStop}%
\bibitem [{\citenamefont {Cai}\ \emph {et~al.}(1994{\natexlab{a}})\citenamefont
  {Cai}, \citenamefont {Bishop}, \citenamefont {Gr{\o}nbech-Jensen},\ and\
  \citenamefont {Malomed}}]{cai1994moving}%
  \BibitemOpen
  \bibfield  {author} {\bibinfo {author} {\bibfnamefont {D.}~\bibnamefont
  {Cai}}, \bibinfo {author} {\bibfnamefont {A.}~\bibnamefont {Bishop}},
  \bibinfo {author} {\bibfnamefont {N.}~\bibnamefont {Gr{\o}nbech-Jensen}}, \
  and\ \bibinfo {author} {\bibfnamefont {B.~A.}\ \bibnamefont {Malomed}},\
  }\bibfield  {title} {\enquote {\bibinfo {title} {Moving solitons in the
  damped ablowitz-ladik model driven by a standing wave},}\ }\href@noop {}
  {\bibfield  {journal} {\bibinfo  {journal} {Physical Review E}\ }\textbf
  {\bibinfo {volume} {50}},\ \bibinfo {pages} {R694} (\bibinfo {year}
  {1994}{\natexlab{a}})}\BibitemShut {NoStop}%
\bibitem [{\citenamefont {Caudrelier}\ and\ \citenamefont
  {Cramp{\'e}}(2019)}]{CAUDRELIER2019114720}%
  \BibitemOpen
  \bibfield  {author} {\bibinfo {author} {\bibfnamefont {V.}~\bibnamefont
  {Caudrelier}}\ and\ \bibinfo {author} {\bibfnamefont {N.}~\bibnamefont
  {Cramp{\'e}}},\ }\bibfield  {title} {\enquote {\bibinfo {title} {New
  integrable boundary conditions for the ablowitz-ladik model: From hamiltonian
  formalism to nonlinear mirror image method},}\ }\href {\doibase
  https://doi.org/10.1016/j.nuclphysb.2019.114720} {\bibfield  {journal}
  {\bibinfo  {journal} {Nuclear Physics B}\ }\textbf {\bibinfo {volume}
  {946}},\ \bibinfo {pages} {114720} (\bibinfo {year} {2019})}\BibitemShut
  {NoStop}%
\bibitem [{\citenamefont {Cai}\ \emph {et~al.}(1994{\natexlab{b}})\citenamefont
  {Cai}, \citenamefont {Bishop},\ and\ \citenamefont
  {Gr\o{}nbech-Jensen}}]{Cai:1994}%
  \BibitemOpen
  \bibfield  {author} {\bibinfo {author} {\bibfnamefont {D.}~\bibnamefont
  {Cai}}, \bibinfo {author} {\bibfnamefont {A.~R.}\ \bibnamefont {Bishop}}, \
  and\ \bibinfo {author} {\bibfnamefont {N.}~\bibnamefont
  {Gr\o{}nbech-Jensen}},\ }\bibfield  {title} {\enquote {\bibinfo {title}
  {Localized states in discrete nonlinear schr\"odinger equations},}\ }\href
  {\doibase 10.1103/PhysRevLett.72.591} {\bibfield  {journal} {\bibinfo
  {journal} {Phys. Rev. Lett.}\ }\textbf {\bibinfo {volume} {72}},\ \bibinfo
  {pages} {591} (\bibinfo {year} {1994}{\natexlab{b}})}\BibitemShut {NoStop}%
\bibitem [{\citenamefont {Lichtenberg}\ and\ \citenamefont
  {Lieberman}(1992)}]{lichtenberg1992regular}%
  \BibitemOpen
  \bibfield  {author} {\bibinfo {author} {\bibfnamefont {A.~J.}\ \bibnamefont
  {Lichtenberg}}\ and\ \bibinfo {author} {\bibfnamefont {M.~A.}\ \bibnamefont
  {Lieberman}},\ }\bibfield  {title} {\enquote {\bibinfo {title} {Regular and
  {C}haotic {D}ynamics, vol. 38 of},}\ }\href@noop {} {\bibfield  {journal}
  {\bibinfo  {journal} {Applied Mathematical Sciences}\ } (\bibinfo {year}
  {1992})}\BibitemShut {NoStop}%
\bibitem [{fre()}]{freelyDOP853}%
  \BibitemOpen
  \href@noop {} {\ }\bibinfo {note} {Freely available from:
  \url{http://www.unige.ch/~hairer/software.html}}\BibitemShut {NoStop}%
\bibitem [{\citenamefont {Hairer}\ \emph {et~al.}(1993)\citenamefont {Hairer},
  \citenamefont {N{\o}rsett},\ and\ \citenamefont
  {Wanner}}]{hairer1993solving}%
  \BibitemOpen
  \bibfield  {author} {\bibinfo {author} {\bibfnamefont {E.}~\bibnamefont
  {Hairer}}, \bibinfo {author} {\bibfnamefont {S.~P.}\ \bibnamefont
  {N{\o}rsett}}, \ and\ \bibinfo {author} {\bibfnamefont {G.}~\bibnamefont
  {Wanner}},\ }\href@noop {} {\emph {\bibinfo {title} {Solving ordinary
  differential equations I. Nonstiff problems, 2nd edition}}},\ Vol.~\bibinfo
  {volume} {14}\ (\bibinfo  {publisher} {Springer Series in Computational
  Mathematics},\ \bibinfo {year} {1993})\BibitemShut {NoStop}%
\bibitem [{\citenamefont {Danieli}\ \emph
  {et~al.}(2019{\natexlab{b}})\citenamefont {Danieli}, \citenamefont {Manda},
  \citenamefont {Mithun},\ and\ \citenamefont {Skokos}}]{mine-01-03-447}%
  \BibitemOpen
  \bibfield  {author} {\bibinfo {author} {\bibfnamefont {C.}~\bibnamefont
  {Danieli}}, \bibinfo {author} {\bibfnamefont {B.~M.}\ \bibnamefont {Manda}},
  \bibinfo {author} {\bibfnamefont {T.}~\bibnamefont {Mithun}}, \ and\ \bibinfo
  {author} {\bibfnamefont {C.}~\bibnamefont {Skokos}},\ }\bibfield  {title}
  {\enquote {\bibinfo {title} {Computational efficiency of numerical
  integration methods for the tangent dynamics of many-body hamiltonian systems
  in one and two spatial dimensions},}\ }\href {\doibase
  http://dx.doi.org/10.3934/mine.2019.3.447} {\bibfield  {journal} {\bibinfo
  {journal} {Mathematics in Engineering}\ }\textbf {\bibinfo {volume} {1}},\
  \bibinfo {pages} {447} (\bibinfo {year} {2019}{\natexlab{b}})}\BibitemShut
  {NoStop}%
\bibitem [{\citenamefont {Liu}\ and\ \citenamefont
  {Tegmark}(2021)}]{PhysRevLett.126.180604}%
  \BibitemOpen
  \bibfield  {author} {\bibinfo {author} {\bibfnamefont {Z.}~\bibnamefont
  {Liu}}\ and\ \bibinfo {author} {\bibfnamefont {M.}~\bibnamefont {Tegmark}},\
  }\bibfield  {title} {\enquote {\bibinfo {title} {Machine learning
  conservation laws from trajectories},}\ }\href {\doibase
  10.1103/PhysRevLett.126.180604} {\bibfield  {journal} {\bibinfo  {journal}
  {Phys. Rev. Lett.}\ }\textbf {\bibinfo {volume} {126}},\ \bibinfo {pages}
  {180604} (\bibinfo {year} {2021})}\BibitemShut {NoStop}%
\bibitem [{\citenamefont {Mithun}\ \emph {et~al.}(2018)\citenamefont {Mithun},
  \citenamefont {Kati}, \citenamefont {Danieli},\ and\ \citenamefont
  {Flach}}]{Mithun:2018}%
  \BibitemOpen
  \bibfield  {author} {\bibinfo {author} {\bibfnamefont {T.}~\bibnamefont
  {Mithun}}, \bibinfo {author} {\bibfnamefont {Y.}~\bibnamefont {Kati}},
  \bibinfo {author} {\bibfnamefont {C.}~\bibnamefont {Danieli}}, \ and\
  \bibinfo {author} {\bibfnamefont {S.}~\bibnamefont {Flach}},\ }\bibfield
  {title} {\enquote {\bibinfo {title} {Weakly {N}onergodic {D}ynamics in the
  {G}ross-{P}itaevskii {L}attice},}\ }\href {\doibase
  10.1103/PhysRevLett.120.184101} {\bibfield  {journal} {\bibinfo  {journal}
  {Phys. Rev. Lett.}\ }\textbf {\bibinfo {volume} {120}},\ \bibinfo {pages}
  {184101} (\bibinfo {year} {2018})}\BibitemShut {NoStop}%
\bibitem [{\citenamefont {Echmann}\ and\ \citenamefont
  {Ruelle}(1985)}]{echmann}%
  \BibitemOpen
  \bibfield  {author} {\bibinfo {author} {\bibfnamefont {J.-P.}\ \bibnamefont
  {Echmann}}\ and\ \bibinfo {author} {\bibfnamefont {D.}~\bibnamefont
  {Ruelle}},\ }\bibfield  {title} {\enquote {\bibinfo {title} {Ergodic theory
  of chaos and strange attractors},}\ }\href {\doibase
  10.1103/RevModPhys.57.617} {\bibfield  {journal} {\bibinfo  {journal} {Rev.
  Mod. Phys.}\ }\textbf {\bibinfo {volume} {57}},\ \bibinfo {pages} {617}
  (\bibinfo {year} {1985})}\BibitemShut {NoStop}%
\bibitem [{\citenamefont {Mithun}\ \emph
  {et~al.}(2021{\natexlab{b}})\citenamefont {Mithun}, \citenamefont {Danieli},
  \citenamefont {Fistul}, \citenamefont {Altshuler},\ and\ \citenamefont
  {Flach}}]{PhysRevE.104.014218}%
  \BibitemOpen
  \bibfield  {author} {\bibinfo {author} {\bibfnamefont {T.}~\bibnamefont
  {Mithun}}, \bibinfo {author} {\bibfnamefont {C.}~\bibnamefont {Danieli}},
  \bibinfo {author} {\bibfnamefont {M.~V.}\ \bibnamefont {Fistul}}, \bibinfo
  {author} {\bibfnamefont {B.~L.}\ \bibnamefont {Altshuler}}, \ and\ \bibinfo
  {author} {\bibfnamefont {S.}~\bibnamefont {Flach}},\ }\bibfield  {title}
  {\enquote {\bibinfo {title} {Fragile many-body ergodicity from action
  diffusion},}\ }\href {\doibase 10.1103/PhysRevE.104.014218} {\bibfield
  {journal} {\bibinfo  {journal} {Phys. Rev. E}\ }\textbf {\bibinfo {volume}
  {104}},\ \bibinfo {pages} {014218} (\bibinfo {year}
  {2021}{\natexlab{b}})}\BibitemShut {NoStop}%
\bibitem [{\citenamefont {Liu}\ and\ \citenamefont
  {Tegmark}(2022)}]{PhysRevLett.128.180201}%
  \BibitemOpen
  \bibfield  {author} {\bibinfo {author} {\bibfnamefont {Z.}~\bibnamefont
  {Liu}}\ and\ \bibinfo {author} {\bibfnamefont {M.}~\bibnamefont {Tegmark}},\
  }\bibfield  {title} {\enquote {\bibinfo {title} {Machine learning hidden
  symmetries},}\ }\href {\doibase 10.1103/PhysRevLett.128.180201} {\bibfield
  {journal} {\bibinfo  {journal} {Phys. Rev. Lett.}\ }\textbf {\bibinfo
  {volume} {128}},\ \bibinfo {pages} {180201} (\bibinfo {year}
  {2022})}\BibitemShut {NoStop}%
\bibitem [{\citenamefont {Bondesan}\ and\ \citenamefont
  {Lamacraft}(2019)}]{lamacraft}%
  \BibitemOpen
  \bibfield  {author} {\bibinfo {author} {\bibfnamefont {R.}~\bibnamefont
  {Bondesan}}\ and\ \bibinfo {author} {\bibfnamefont {A.}~\bibnamefont
  {Lamacraft}},\ }\bibfield  {title} {\enquote {\bibinfo {title} {Learning
  symmetries of classical integrable systems},}\ }\href@noop {} {\bibfield
  {journal} {\bibinfo  {journal} {arXiv preprint arXiv:1906.04645}\ } (\bibinfo
  {year} {2019})}\BibitemShut {NoStop}%
\end{thebibliography}%

\end{document}